\documentclass[twocolumn,prb,10pt,aps,superscriptaddress,longbibliography,nofootinbib,floatfix]{revtex4-1}

\usepackage{amsmath}
\usepackage{amssymb}
\usepackage{amsthm}
\usepackage{amsfonts}

\usepackage[all]{xy}
\usepackage{graphicx}
\usepackage{braket}

\usepackage{color}
\definecolor{darkblue}{rgb}{0.,0.,0.4}
\definecolor{darkred}{rgb}{0.5,0.,0.}
\usepackage[pdftex,colorlinks=true,linkcolor=darkblue,citecolor=blue,urlcolor=darkred]{hyperref}

\DeclareMathOperator{\Tr}{Tr}
\DeclareMathOperator{\Cor}{Cor}
\newcommand{\cG}{\mathcal{G}}
\newcommand{\mb}[1]{\mathbb{#1}}

\begin{document}

\title{Spurious Long-range Entanglement and Replica Correlation Length}

\author{Liujun Zou}
\affiliation{Department of Physics, Harvard University, Cambridge, MA}
\author{Jeongwan Haah}
\affiliation{Department of Physics, Massachusetts Institute of Technology, Cambridge, MA}

\begin{abstract}
Topological entanglement entropy has been regarded as a smoking-gun signature of topological order in two dimensions,
capturing the total quantum dimension of the topological particle content.
An extrapolation method on cylinders has been used frequently to measure the topological entanglement entropy.
Here, we show that a class of short-range entangled 2D states, when put on an infinite cylinder of circumference $L$,
exhibits the entanglement R\'enyi entropy of any integer index $\alpha \ge 2$ that obeys
$S_\alpha = a L - \gamma$ where $a, \gamma > 0 $.
Under the extrapolation method,
the subleading term $\gamma$ would be identified as the topological entanglement entropy,
which is spurious.
A nonzero $\gamma$ is always present if the 2D state reduces
to a certain symmetry-protected topological 1D state,
upon disentangling spins that are far from the entanglement cut.
The internal symmetry that stabilizes $\gamma > 0$
is not necessarily a symmetry of the 2D state,
but should be present after the disentangling reduction.
If the symmetry is absent, $\gamma$ decays exponentially in $L$
with a characteristic length, termed as a replica correlation length,
which can be {\em arbitrarily large} compared to the two-point correlation length of the 2D state.
We propose a simple numerical procedure to measure the replica correlation length
through replica correlation functions.
We also calculate the replica correlation functions for representative wave functions
of abelian discrete gauge theories and the double semion theory in 2D,
to show that they decay abruptly to zero.
This supports a conjecture that the replica correlation length being small
implies that the subleading term from the extrapolation method determines the total quantum dimension.
\end{abstract}

\date{23 July 2016}

\maketitle

\section{Introduction} \label{sec:introduction}

Topologically ordered states are nontrivial gapped states,
which are beyond Landau's symmetry breaking paradigm.\cite{Wen2004Book}
Prototypical examples include gapped spin liquid states and fractional quantum Hall states.
These states exhibit robust ground state degeneracy
depending on the topology of the system,
and fractionalized excitations.\cite{Wen1990,Oshikawa2006}
Even more intriguing is its potential of
being a fault-tolerant quantum information processing
platform.\cite{Kitaev2003Fault-tolerant,NayakEtAl2008tqc}
Recently, long-range entanglement has been appreciated
to understand the topological order,
as topologically nontrivial states cannot be connected to a product state by local unitary
transformations.\cite{BravyiHastingsVerstraete2006generation,
Chen2010,Hastings2010Locality,Haah2014invariant}

Detecting topological order, however, has still been a challenge.
The ground state degeneracy and the fractional quantum numbers of the excitations are difficult to measure
even in the numerical calculations, let alone experimental situations.
Instead, the so-called topological entanglement entropy
(TEE)\cite{HammaIonicioiuZanardi2005,
AliosciaHammaZanardi2005Ground,
KitaevPreskill2006Topological,
LevinWen2006Detecting,
FlammiaHammaHughesWen2009TEE,
HalaszHamma2012Probing}
is being recognized as an important quantity especially in numerics,
for the purpose of distinguishing topologically ordered states from topologically trivial states.

It is believed that the bipartite entanglement entropy
of the ground state of a gapped system in two spatial dimensions
obeys an ``area'' law with a constant sub-leading correction.
Specifically, the entanglement entropy for a disk of circumference $L$ is given by
\begin{equation}
S=\alpha L-\gamma+\cdots
 \label{eq:arealaw}
\end{equation}
where $\alpha$ is a model-specific non-universal coefficient,
$\gamma$ is a sub-leading correction,
and the ellipses represent terms that vanish in the large $L$ limit.
This constant correction $\gamma$ is the universal TEE of the state.
It is shown that $\gamma = \log \mathcal D$,
where $\mathcal D$ is the so-called total quantum dimension of the system,
determined by the anyon content of the topological order that the state represents.
Roughly speaking, $\mathcal D$ counts the types of fractionalized particles.
Since $\mathcal D > 1$ implies the state supports fractionalized excitations,
$\gamma$ is regarded as a smoking-gun signature of 2D topological order.

\begin{figure}[bt]
  \centering
  \includegraphics[width=0.4\textwidth]{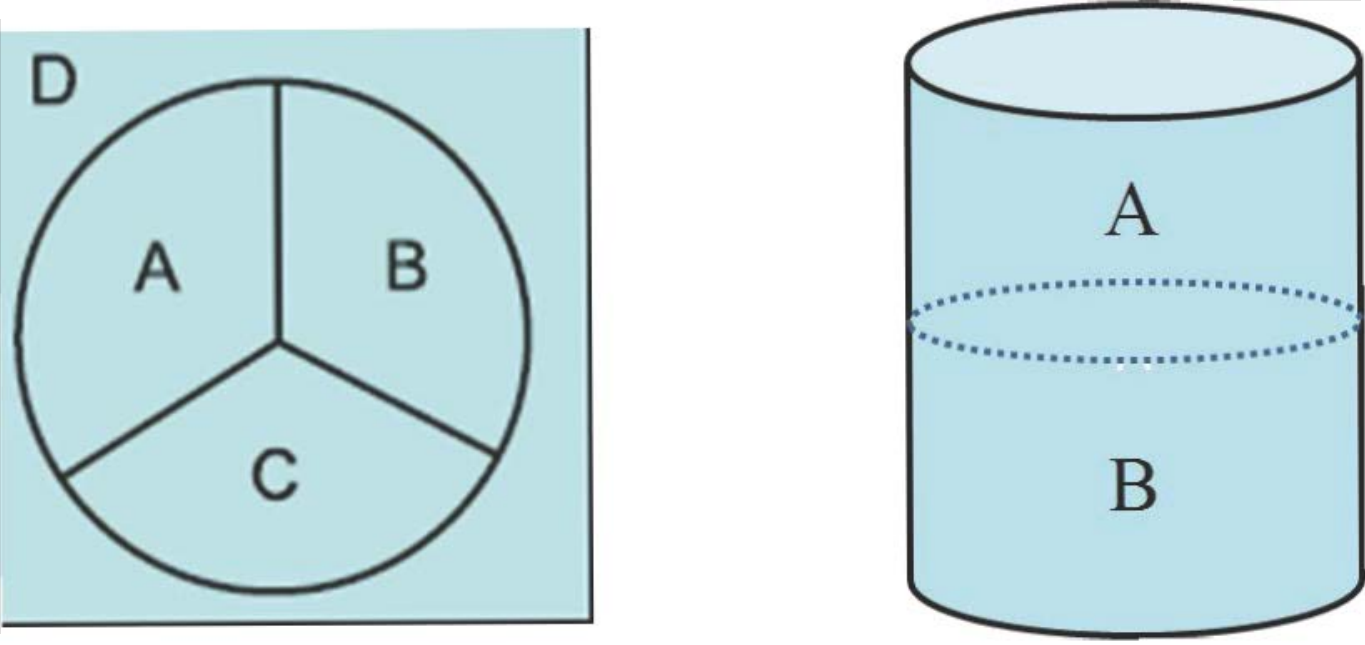}
  \caption{Two methods of extracting TEE.
  Left: Kitaev-Preskill prescription divides the system into four parts and extract TEE by using \eqref{eq:Kitaev-Preskill}.
  Right: DMRG calculations put the system on an infinite cylinder and divide the system into two parts,
  then calculate the entanglement entropy $S(L)$ between the two parts for different circumferences $L$ of the cylinder.
  Fitting the results into \eqref{eq:arealaw}, TEE is identified as $-S(L=0)$.}
  \label{fig:extract-TEE}
\end{figure}

Remark that the definition of $\gamma$ through Eq.~\eqref{eq:arealaw} is inherently ambiguous;
it depends on fine details of a regularization scheme for the calculation of $S$.
On a lattice, the circumference $L$ and therefore the subleading term $\gamma$
vary according to how one counts the number of sites along the boundary of the disk;
for example,
the circumference of a rectangle that encloses $L \times L$ sites
may be counted as $(L+2)^2 - L^2 = 4L+4$ or $L^2 - (L-2)^2 = 4L-4$.
To resolve this ambiguity by eliminating the boundary term,
Ref.~\onlinecite{KitaevPreskill2006Topological,LevinWen2006Detecting}
take a linear combination of entanglement entropies for various regions.
Concretely, Ref.~\onlinecite{KitaevPreskill2006Topological}
proposes the following combination.
\begin{equation} \label{eq:Kitaev-Preskill}
\gamma=S_{AB}+S_{BC}+S_{CA}-S_A-S_B-S_C-S_{ABC}
\end{equation}
where the subscripts refer to the regions specified in Fig.~\ref{fig:extract-TEE}.
In fact, it is this linear combination that enables one to argue
that $\gamma$ is a robust quantity under small changes in the Hamiltonian and the region sizes;
the combination contains an equal number of terms of opposite signs for each subregion
so that a small change in any subregion may be canceled overall.
We will refer to this proposal as the {\em Kitaev-Preskill prescription} hereafter. Note that it is not too important at long distances whether
one uses von Neumann entropy or
R\'enyi entropy.\cite{FlammiaHammaHughesWen2009TEE,HalaszHamma2012Probing}

Despite of its conceptual importance,
the Kitaev-Preskill prescription of extracting TEE
is not of great practical use
because it requires each partition be much larger than the correlation length of the system.\cite{PapanikolaouRamanFradkin2007Topological}
This is very challenging especially in density-matrix-renormalization-group (DMRG) methods.
Alternatively, by exploiting the fact that the DMRG algorithm
systematically produces minimally entangled states,
it is proposed that one can simply put the system on infinite cylinders
with various circumferences $L$ (see Fig.~\ref{fig:extract-TEE}),
and extrapolate the data using \eqref{eq:arealaw} to read off $\gamma$.\cite{JiangWangBalents2012}
The mentioned ambiguity of defining the circumference of a disk on lattices,
does not apply here since the circumference
of a cylinder is simply well-defined.
We will refer to this method as the {\em cylinder extrapolation method}.
An important advantage of this method is that
one can regard the region size as large as $L$,
rather than some small fraction of $L$,
and hence can expect finite size effects to be very small.
This method might be useful in light of recent experimental developments
as there are proposals to measure entanglement entropies,
and experiments have already been performed on simple cold atom systems.\cite{PichlerBonnesDaleyEtAl2013,Islam2015}

However, the cylinder extrapolation method seems to yield inconsistent results.
By applying the cylinder extrapolation method to the $J_1$-$J_2$ antiferromagnetic
Heisenberg model on a square lattice,
it is found that $S(L=0) \simeq -\ln 2$ in a certain parameter regime,\cite{JiangYaoBalents2012square}
and thus the ground state is identified as a topologically ordered spin liquid.
However, this result was later objected by another DMRG study
from an independent group, revealing a plaquette valence-bond order.\cite{GongZhuShengEtAl2014square}
Furthermore, Ref.~\onlinecite{GongShengMotrunichEtAl2013honeycomb}
studies the Heisenberg model on the honeycomb lattice,
and reports $S(L=0) \simeq -\ln 2$ by the cylinder extrapolation method,
but observes a plaquette valence-bond order, which leads to
a suspicion of finite size effects on $S(L=0) < 0$.
These numerical results question the validity of the cylinder extrapolation method.

In this paper, we point out one scenario
under which the cylinder extrapolation method can be proven to be {\em invalid}.
While we do not attempt to address specific reasons
behind the discrepancy between Ref.~\onlinecite{JiangYaoBalents2012square} and
\onlinecite{GongZhuShengEtAl2014square},
we will give a sufficient condition for topologically {\em trivial} 2D states,
under which the cylinder extrapolation method must give a nonvanishing sub-leading term
for any R\'enyi entropy calculations,
leading to a spurious TEE.

We start with a general observation that
when a two-dimensional state is topologically trivial,
the entanglement computation in Fig.~\ref{fig:extract-TEE}(b)
reduces to that of a one-dimensional state
with respect to an extensive bipartition.
We show that whenever the derived one-dimensional state exhibits
a symmetry-protected topological (SPT) order under a product group,
which we define precisely below,
then the cylinder extrapolation method must output a nonzero sub-leading term.

Under generic perturbations that break the symmetry of 
the reduced one-dimensional states in our examples,
the sub-leading term is suppressed exponentially in the system size.
Furthermore, if one applies the Kitaev-Preskill prescription in the bulk,
one still obtains a value that is consistent with the total quantum dimension
of the underlying topological particle content.
Hence, it is improper to say that the notion of topological entanglement entropy
is invalidated.
Rather, our examples make it clear that one has to be
careful in interpreting results from the cylinder extrapolation method.

To decide when the results from the cylinder extrapolation method
can be trusted,
we consider a length scale,
termed {\em replica correlation length} $\xi_\alpha$.
The ratio $L/\xi_\alpha$ determines the magnitude
the sub-leading term in the cylinder extrapolation method.
This replica correlation length
may be arbitrarily large compared to the usual correlation length of the 2D state.
The usual correlation length of a state $\rho$
is the decay rate of (the connected part of)
a two-point function $\Tr(\rho O O')$
as a function of the distance between local operators $O$ and $O'$.
In contrast, the replica correlation length
is the decay rate of a two-point replica correlation function
$\Tr(\rho \mathcal O_1 O_1' \rho O_2 O_2' )/\Tr(\rho^2)$
where the unprimed operators and primed operators are far separated.

At the first glance on its definition,
the replica correlation length may seem difficult to calculate.
We propose a relatively simple way of measuring the replica correlation function in numerics,
as a natural extension of the measurement of R\'enyi entropies using swap
operations.\cite{HastingsGonzalezKallinMelko2010}
We find that a 2D cluster state (graph state\cite{HeinEisertBriegel2004graph}),
which is topologically trivial,
has an infinite replica correlation length,
while certain representative wave functions of the $\mathbb Z_N$ gauge theory
and the double-semion theory have
replica correlation length zero. We conjecture that the $\gamma$ value from the cylinder extrapolation method
is given by the total quantum dimension
whenever the replica correlation length is small.

Some of our examples can be adapted to three or higher dimensions\cite{GroverTurnerVishwanath2011TEE}
and give similar effects.
It would also be interesting to consider thermal states.\cite{CastelnovoChamon2007Entanglement,
CastelnovoChamon2008Topological,
Hastings2011warmTQO}
We will comment on these in the discussion section.

The rest of the paper is organized as follows.
In Sec.~\ref{sec:2dcluster}, we study the 2D cluster state on a triangular lattice,
an exactly soluble model of topologically trivial 2D states,
and show that the sub-leading term of the entanglement R\'enyi entropy
calculated by the cylinder extrapolation method is nonzero.
This example illustrates important points and serves
as a warm-up for the more general consideration.
In Sec.~\ref{sec:generalconstruction},
we study the R\'enyi entropies of a class of topologically trivial 2D states
and show that the sub-leading term is strictly negative,
if it is calculated using the cylinder extrapolation method.
In particular, we will attribute the nonzero sub-leading term
to a SPT order under a product group of the derived one-dimensional state.
In Sec.~\ref{sec:RCL},
we consider generic non-symmetric states and
introduce the replica correlation length,
that is responsible for the non-vanishing sub-leading term of entanglement entropy.
We calculate the replica correlation length
to show that it is infinite for the 2D cluster state on triangular lattice,
while it is zero for some ideal wave functions of the
$\mathbb{Z}_N$ gauge theories and double-semion theory.
We conclude in Sec.~\ref{sec:discussion} with discussion on
higher dimensions and thermal states.
Appendices include further considerations.
In App.~\ref{app:intertwinedCluster},
we provide an example where the sub-leading term calculated by the cylinder extrapolation method
oscillates with the system size.
In App.~\ref{app:othersymmetries},
we discuss lattice symmetries and time reversal symmetries,
and find that these symmetries are not responsible for
a robust non-vanishing sub-leading term by the cylinder extrapolation method.
App.~\ref{app:thermal}
contains calculation of the mutual information
of the cluster state at nonzero temperature.

\section{Example: 2D cluster state on a triangular lattice}
\label{sec:2dcluster}

\begin{figure}[t]
\centering
\includegraphics[width=0.15\textwidth]{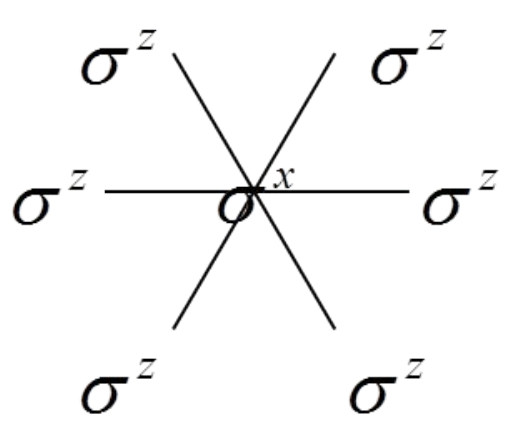}\\
\caption{A 7-spin interaction in the summation of Eq.~\eqref{eq:clusterHamiltonian},
  which is a product of the $\sigma^x$ operator on the center site
  and the $\sigma^z$ operators on the sites that surround it.}
\label{fig:7-spin}
\end{figure}

In this section we study an example of topologically trivial 2D state
that nevertheless exhibits nonvanishing sub-leading term of entanglement entropy
under the cylinder extrapolation method.
We start with a triangular lattice with one spin-1/2 (qubit) $\{ \ket 0, \ket 1 \}$
per lattice site,
governed by a Hamiltonian
\begin{equation}
 H_0 = - \sum_j \sigma_j^x
\end{equation}
The ground state $\ket{\psi_0}$ of $H_0$ is a product state and there is no entanglement.
Clearly, this state is topologically trivial.
Next, we apply to $\ket{\psi_0}$ a layer of local unitary transformations $U_{jk}$
for each pair $\langle j k \rangle$ of nearest neighbor qubits.
The two-qubit unitary $U = U_{jk} = U_{kj}$
is most conveniently defined in a basis where $\sigma^z$ is diagonal:
\begin{equation}
 U \ket{ a} \ket{ b} =
 \begin{cases}
  - \ket{1}\ket{1} &\text{if } a=b=1,\\
  +\ket{a}\ket{b} &\text{otherwise.}
 \end{cases}
 \label{eq:defnCPHASE}
\end{equation}
Since $U_{jk}$ are simultaneously diagonal in a basis,
they commute with each other
\begin{equation}
 U_{jk} U_{j'k'} = U_{j'k'} U_{jk}.
\end{equation}
Hence, there is no ambiguity in the formula
\begin{equation}
 \mathbf U = \prod_{\langle jk \rangle } U_{jk},
 \label{eq:ProdCPhaseU}
\end{equation}
and we define a state $\ket \psi = \mathbf U \ket{\psi_0}$,
which is called the cluster state.
Using the identity
\begin{equation}
 U_{jk} (\sigma^x_j \otimes I) U_{jk}^\dagger = \sigma^x_j \otimes \sigma^z_k,
\end{equation}
we see that the cluster state is a ground state of a Hamiltonian
\begin{align} \label{eq:clusterHamiltonian}
 H = \mathbf{U} H_0 \mathbf{U}^\dagger = - \sum_j \left( \sigma^x_j \prod_{k: \langle j k \rangle } \sigma_{k}^z \right).
\end{align}
Graphically, each term in the summation of the new Hamiltonian \eqref{eq:clusterHamiltonian}
is the 7-spin interaction as shown in Fig.~\ref{fig:7-spin}.

\subsection{Entanglement entropy on a cylinder}

\begin{figure}[b]
 \includegraphics[width=.4\textwidth]{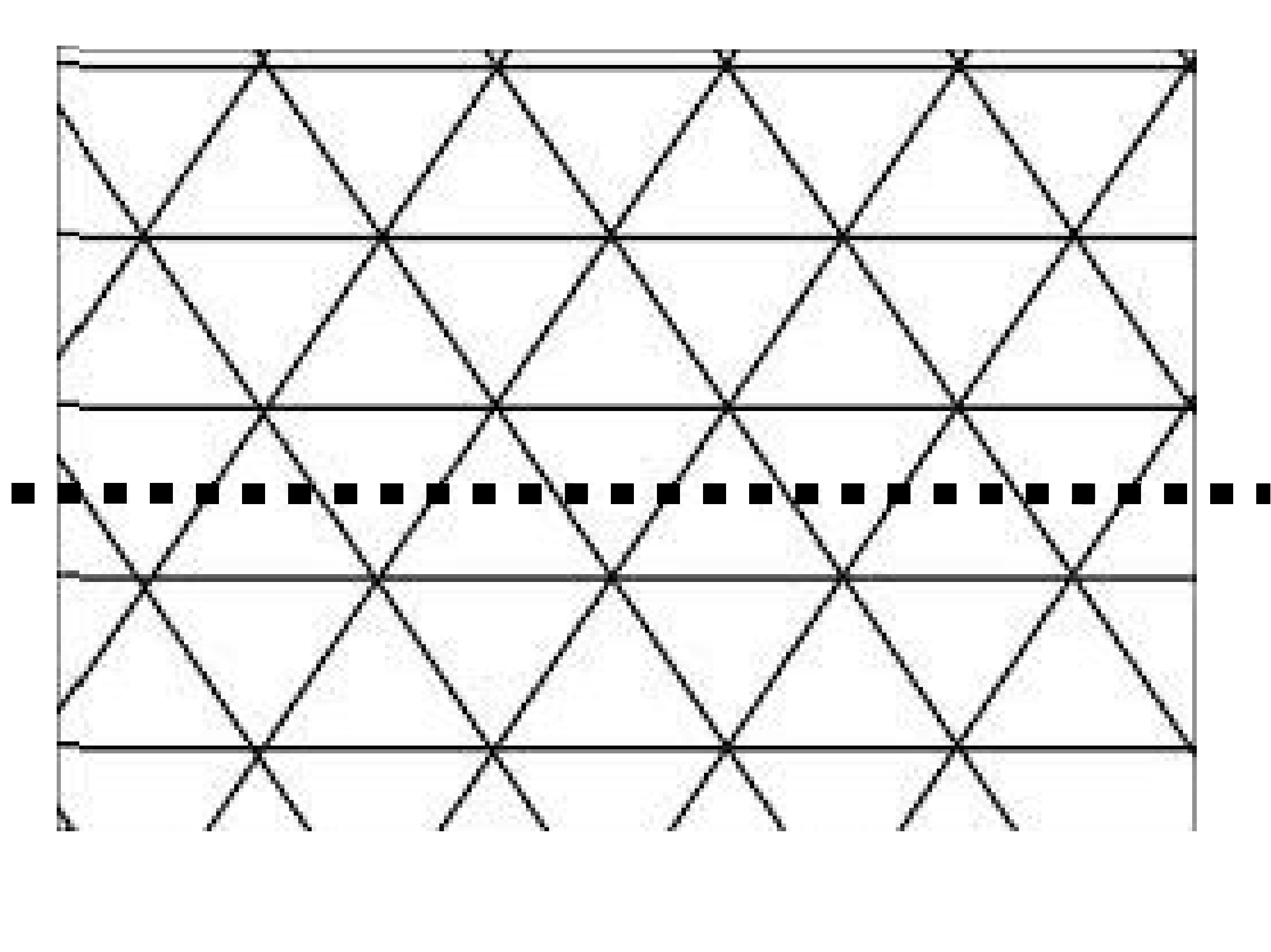}
 \caption{Triangular lattice with an entanglement cut parallel to one of the sides of a triangle.
 The upper region is $A$, and the lower is $B$.}
 \label{fig:triangular-lattice}
\end{figure}

Since we obtain the cluster state from a product state by a small depth quantum circuit,
the cluster state is also topologically trivial.
One may expect its bipartite entanglement entropy has a vanishing sub-leading term.
This is indeed the case if we use the Kitaev-Preskill prescription
to extract the sub-leading term.
Now let us examine it using the cylinder extrapolation method.
We put our state on an infinite cylinder
by imposing a periodic boundary condition
along one of three directions parallel to any side of a triangle.
If the circumference is $L$,
the number of bond cuts is $2L$.
We will compute the entanglement entropy
between the two sides $A, B$ divided by this circumference
(see Fig. ~\ref{fig:triangular-lattice}).

\begin{figure}[b]
  \centering
  \includegraphics[width=0.4\textwidth]{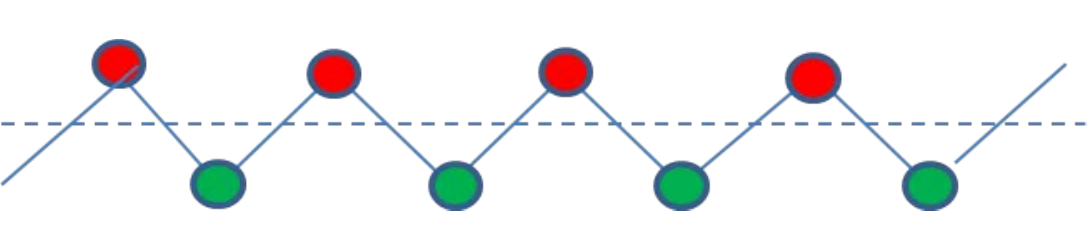}
  \caption{The reduced 1D chain of the 2D cluster state.
  The red circles represent qubits in region A and the green circles represent qubits in region B.}
  \label{fig:zig-zag}
\end{figure}

Even though we started with a 2D state,
the entropy computation reduces to that of a 1D chain with an extensive bipartition.
To see this, recall that the entanglement entropy is invariant
under any unitary that acts exclusively on either side of the bipartition:
\begin{equation}
 S = S( \rho_A ) = S( U_A \rho_A U_A^\dagger ) = S( \rho_B ) = S(U_B \rho_B U_B^\dag )
\end{equation}
where the subscript $A$ or $B$
denotes the region on which the operator is supported.
In particular, we can choose $U_A$ to be the product of all $U_{jk}$
where the {\em edge} $\langle jk \rangle$ belongs to $A$.
Since $U_{jk}^2 = I$, this amounts to disentangling $\ket \psi$ on the region $A$.
A similar disentangling unitary can be applied on $B$.
What is left is a zig-zag 1D chain
that straddles two regions along the cut,
and some completely disentangled qubits in the product state.
See Fig.~\ref{fig:zig-zag}.
It remains to compute the entanglement entropy of the 1D chain $\ket{ \psi_1}$,
which is a ground state of
\begin{align}
 H_1 &= - \sum_{j=1}^L \sigma^x_{j,A}  \sigma^z_{j-1,B} \sigma^z_{j,B} - \sum_{j=1}^L \sigma^x_{j,B} \sigma^z_{j,A} \sigma^z_{j+1,A} \nonumber \\
 &= -\sum_{k=1}^{2L} \sigma^z_{k-1} \sigma^x_{k} \sigma^z_{k+1} \label{eq:1dClusterH}
\end{align}
where a periodic boundary condition is imposed,
i.e., $j=L+1$ site is equal to $j=1$ site.
Note that since the Hamiltonian is commuting and the ground state is non-degenerate,
any correlation function is identically zero
beyond distance $2$, and hence the correlation length vanishes.

One might guess that the entanglement entropy of $\ket{\psi_1}$
is just proportional to the number of bond cuts,
because the entangling unitary $U_{jk}$ makes the product state $\ket{+}_j \ket{+}_k$
into a maximally entangled state,
where $\ket + = (\ket 0 + \ket 1 )/\sqrt 2$.
However, a more careful and direct computation reveals that
\begin{equation}
S(A) = (\log 2) L - \log 2.
\label{eq:ZXZ-entropy}
\end{equation}
This computation can be done by exploiting the fact that the Hamiltonian
$H_1$ consists of commuting tensor products of Pauli matrices.
It is known that the eigenvalue spectrum
of the reduced density matrix $\rho_A$ for any set of qubits $A$
consists of a single nonzero value (flat entanglement spectrum),
and the number of nonzero eigenvalues is always $2^k$ for some integer $k \ge 0$.
The exponent $k$ depends only on
the number of operators $P$ that are products of Pauli matrices
and stabilize the state, $P \ket \psi = \ket \psi$;
more precisely,
$k$ equals the number of qubits in the region $A$
minus the logarithm of the order of the stabilizer group $\cG_A$
supported on $A$,
\begin{align}
 k = |A| - \log_2 |\cG_A|.
\end{align}
See Section~\ref{sec:FlatES} for a simple proof.
In the present example, there are $L$ qubits in $A$,
and there is a single non-identity stabilizer $\prod_{j=1}^L \sigma^x_{j,A}$
supported on $A$, and hence Eq.~\eqref{eq:ZXZ-entropy} follows.
From Eq.~\eqref{eq:ZXZ-entropy},
one may mistakenly conclude that $\gamma=\log 2$ from the cylinder extrapolation method
and that the original 2D state is topologically ordered.

In the next section, we explain
why such a topologically trivial state
can give rise to a nonzero sub-leading term
in the entanglement entropy calculated by the cylinder extrapolation method.
We will attribute the nonzero sub-leading correction term $\gamma = \log2$
to the nontriviality of the state $\ket{\psi_1}$
as a SPT order under symmetry $\mb{Z}_2 \times \mb{Z}_2$,
where the first $\mb{Z}_2$ factor acts by $\sigma^x$
on the red side of the entanglement cut,
and the second $\mb{Z}_2$ factor by $\sigma^x$ on the green side
of Fig.~\ref{fig:zig-zag}.

\subsection{Reduction to 1D state and its symmetries}

Before proceeding, we remark that for any state constructed from a product state by a small-depth quantum circuit
the entanglement entropy calculation reduces to that of a 1D chain with an extensive partition.
Therefore, our consideration of 1D chains is appropriate and general
to study the bipartite entanglement entropy of 2D topologically trivial states.
The proof of this remark is simple. One can remove all entangling unitaries except for those near the cut
without changing the entanglement.
The unitaries that cannot be removed are supported on a strip
whose width is proportional to the number of layers in the quantum circuit.
Hence, one finally arrives at a quasi-1D system with a bipartition {\em along} the extended direction.
Note that this simple argument proves the area law of entanglement entropy for such 2D states.

We also emphasize that the symmetry of the resulting 1D state is {\em not necessarily} the symmetry of the original 2D state.
The disentangling deformations have no reason to obey any symmetry of the 2D state;
their role is purely to transform the 2D state to a 1D chain immersed in a product state background.
This means that even if the resulting 1D state is symmetric or close to symmetric,
this symmetry is not necessarily visible in the 2D state.
As an example, one can consider the deformation of the 2D triangular lattice cluster state
in the form discussed in Sec.~\ref{sec:RCL}.
One could deform the bonds arbitrarily except those that are crossed by the entanglement cut
such that there is no on-site symmetry.
After the disentangling transformations for the entanglement entropy evaluation,
one can find $\mathbb Z_2 \times \mathbb Z_2$ symmetry of the 1D chain.

Because the symmetry of the reduced 1D state
is obscure from the {\it original} 2D system,
we will introduce  a replica correlation length in Sec.~\ref{sec:RCL},~\ref{sec:RCF} below.
It is distinct from the usual correlation length
but can be checked directly for the original 2D system
without investigating a hidden 1D state.

\subsection{Bravyi's example}
\label{sec:bravyi}

\begin{figure}[h]
  \centering
  \includegraphics[width=0.2\textwidth]{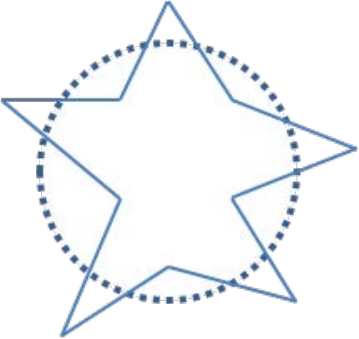}
  \caption{Bravyi's example.
  There is a qubit on each vertex of the zig-zag chain,
  and the dashed line represents an entanglement cut that divides the chain into two halves.
  If the chain is in the cluster state of Eq.~\eqref{eq:1dClusterH},
  the entanglement entropy is equal to $\frac12 \#$(bond cuts) - 1 in the units of $\log 2$.
  }
  \label{fig:Bravyi}
\end{figure}

Bravyi has considered the cluster state on a circular zig-zag chain on a plane,
where an entanglement cut is chosen such that exactly half of the
chain is inside the cut.\cite{Bravyi2008}
The rest of the plane is assumed to be occupied by
qubits in a trivial product state.
Dividing the disk into three circular sectors
(see Figure ~\ref{fig:Bravyi}),
and taking the Kitaev-Preskill combination,
one will find that $\log 2$ remains.
This is, as far as we know, the first example
in which Kitaev-Preskill combination can be nonzero
for a topologically trivial state.
Levin-Wen combination gives no different answer.
The state is highly inhomogeneous, and the partition must be introduced very carefully.
In contrast, our example is manifestly translation-invariant,
and more relevant to current DMRG methods.
Bravyi's example and ours are of course closely related:
If we cap off one end of the cylinder to turn it into a topological plain,
then our zig-zag chain and partition becomes those of Bravyi's.

\section{A sufficient condition for a nonzero sub-leading term of entanglement entropy under the cylinder extrapolation method}
\label{sec:generalconstruction}

In this section,
we will provide a sufficient condition for a class of topologically trivial 2D states
under which these states nevertheless give a nonzero sub-leading term of entanglement entropy
under the cylinder extrapolation method.
Since the bipartite entanglement entropy of any topologically trivial 2D states
is identical to that of a reduced 1D chain under an extensive bipartition,
this amounts to find a condition for such 1D chains
to have a nonvanishing sub-leading term in the entanglement entropy.

We will show for any nontrivial 1D SPT under a product group symmetry $G_1 \times G_2$ (defined more precisely below),
there must be a nonzero negative sub-leading term to the entanglement entropy
with respect to a bipartition where the partition $i = 1,2$
includes all degrees of freedom acted on by $G_i$.
In Fig.~\ref{fig:zig-zag}, for example, $G_1$ acts on the red sites and $G_2$ acts on the green sites.
Here the entanglement entropy is measured by the R\'enyi entropy
\begin{equation}
 S_\alpha(\rho) = \frac{1}{1-\alpha} \log \Tr( \rho^\alpha ),
\end{equation}
and our state is assumed to have a matrix-product state (MPS) representation
with a finite virtual bond dimension.
We restrict ourselves to R\'enyi entropy of integer indices $\alpha = 2,3,4,\ldots$.
The von Neumann entropy ($S_{\alpha \to 1}$) will not be treated explicitly.

\subsection{Nontrivial SPT}

First, we specify what a nontrivial SPT under a product group is.
Recall that 1D SPTs are classified by the second group cohomology $H^2(G; U(1))$,
which enumerates all equivalence classes of factor systems
\begin{equation}
 \omega : G \times G \to U(1)
\end{equation}
obeying the cocycle condition
\begin{equation}
 \frac{\omega(b,c)\omega(a,bc)}{\omega(ab,c)\omega(a,b)} = 1 \quad \text{for all } a,b,c \in G
\end{equation}
up to {\em exact cycles} defined by $\delta \lambda(a,b) := \lambda(a)\lambda(b)/\lambda(ab)$
for some function $\lambda:G \to U(1)$.\cite{SchuchPerez-GarciaCirac2010Classifying,Chen2011}

Now, suppose the symmetry group is $G = G_1 \times G_2$,
and each component acts on different physical qudits.
Graphically, $G_1$ acts on the red sites in Fig.~\ref{fig:zig-zag} and $G_2$ acts on the green sites there.
We say {\em an SPT state under a product group is nontrivial} if its associated factor system $\omega$ admits
\begin{equation}
 \Omega := \frac{\omega(a,b)}{\omega(b,a)} \neq 1 \quad \text{ for some } a \in G_1, b \in G_2.
\end{equation}
Note that $\Omega$ is independent of the cohomology representative $\omega$ since multiplying $\omega$ by an exact cycle does not change $\Omega$.
Since any factor system gives rise to a projective representation $V$,
$\Omega \neq 1$ means that a commuting pair of elements $a,b$ of $G$
are represented by a pair of non-commuting unitaries:
\begin{equation}
 V_a V_b = \frac{1}{\omega(a,b)} V_{ab} = \frac{1}{\omega(a,b)} V_{ba}
 = \frac{\omega(b,a)}{\omega(a,b)} V_b V_a \neq V_b V_a
\end{equation}
In the matrix product state (MPS) representation,
this projective representation appears
in how the symmetry action is implemented on the virtual level.
That is, if the local tensor for a translation-invariant SPT state is given by
\begin{equation}
\mathbb M = \sum_{p,v,w} M^{(p)}_{vw} \ket{p} \otimes \ket {v} \bra{w} ,
\end{equation}
the state is acted on by the symmetry such that
\begin{equation}
 \sum_{p'} (U_g)^p _{~p'} M^{(p')} = \eta_g V_g^\dagger M^{(p)} V_g
 \label{eq:symmetryLifting}
\end{equation}
for any $g \in G$, where $\eta_g$ is a phase factor.\cite{PerezGarciaWolfSanzEtAl2008}
Thus, if the SPT is nontrivial,
the commuting symmetry actions are lifted to non-commuting unitaries on the virtual level.

Eq.~\eqref{eq:symmetryLifting} is most conveniently expressed in diagrams.
If $U$ is a matrix representing the symmetry action of $G_1$, it is
\begin{equation}
\raisebox{-3ex}{\includegraphics[height=8ex]{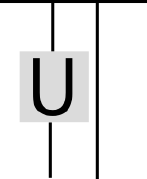}}  =
\raisebox{-3ex}{\includegraphics[height=10ex]{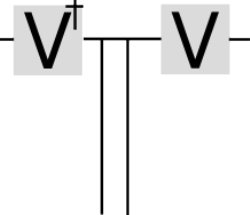}}
\label{eq:diagSymLifting}
\end{equation}
where we have drawn two lines for the physical qudit,
to emphasize that we assumed two components of the symmetry group
act on distinct physical qudits.
We have omitted the phase factor $\eta_g$.
If $U$ were representing $G_2$, the box of $U$ on the left-hand side
would have been on the second vertical line.
Typically in literature, a box is inserted at the intersection of lines to signify a tensor $\mathbb M$,
but we omitted it.

\subsection{Transfer matrix}

\begin{figure}
 \includegraphics[width=.4\textwidth]{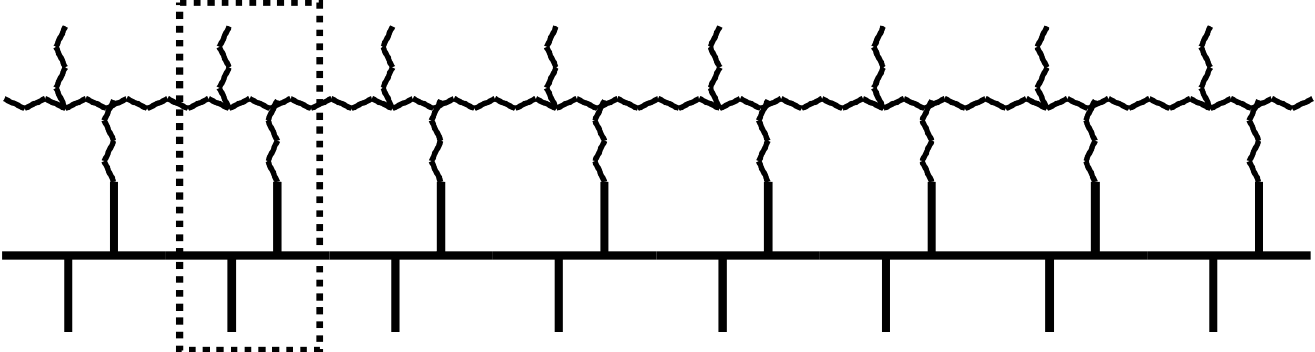}
 \begin{center}(a)\end{center}
 \includegraphics[width=.4\textwidth]{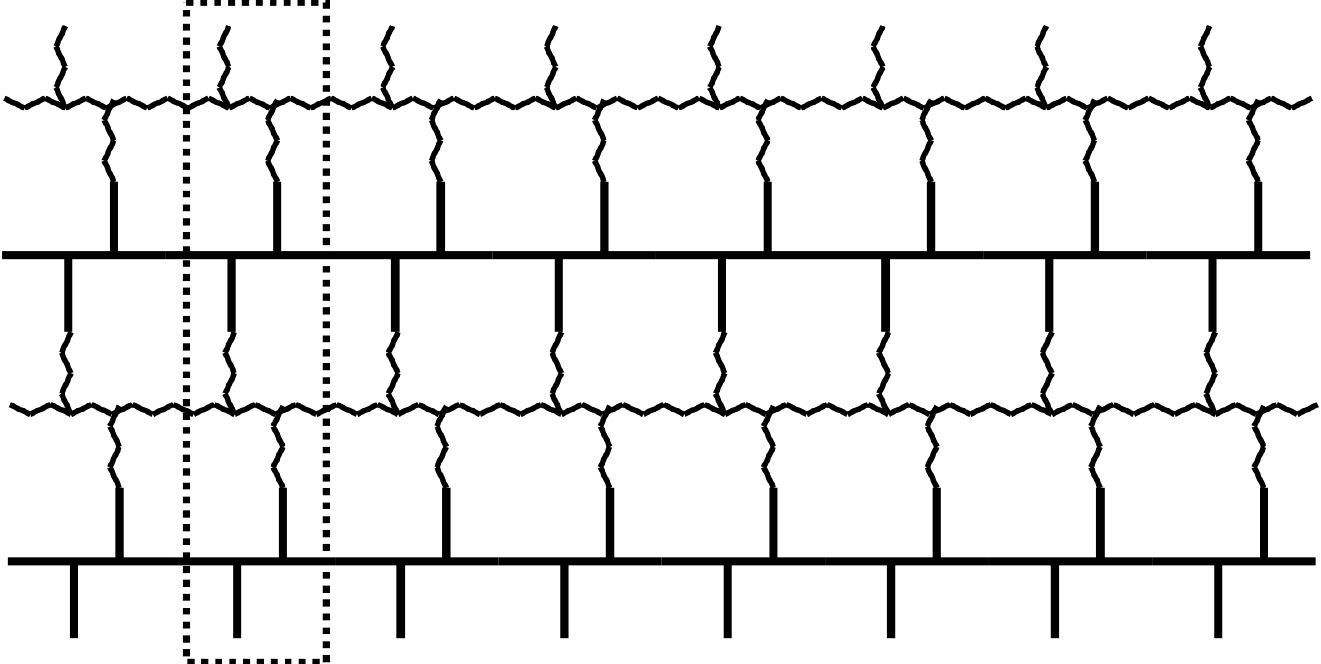}
 \begin{center}(b)\end{center}
 \caption{(a) $\rho$ obtained by tracing out one physical qudits for every site.
 (b) $\rho^2$. Connected bonds are contracted. Wiggling lines represent complex conjugation.
The horizontal virtual bonds are contracted due to the periodic boundary condition, which is not drawn.
$\Tr(\rho)$ and $\Tr(\rho^2)$ are computed by contracting the upper vertical wiggling bonds with the lower straight ones.
For this purpose, it is enough and more efficient to consider the transfer matrix designated by the dotted rectangle;
see Eq.~\eqref{eq:trace-is-trace}.}
 \label{fig:rhos}
\end{figure}

Consider a 1D chain on a ring.
Suppose there are $L$ physical sites with two physical qudits per site, and the two qudits on each site will be acted on by $G_1$ and $G_2$, respectively.
Tracing out one qudit for every site,
we have a reduced density matrix $\rho$.
In a diagram, $\rho$ is depicted as Figure~\ref{fig:rhos}(a).
An integral power of the reduced density matrix can also be represented by a diagram.
For example, $\rho^2$ is depicted in Figure~\ref{fig:rhos}(b).

\begin{figure}[b]
\begin{center}
\includegraphics[width=.47\textwidth]{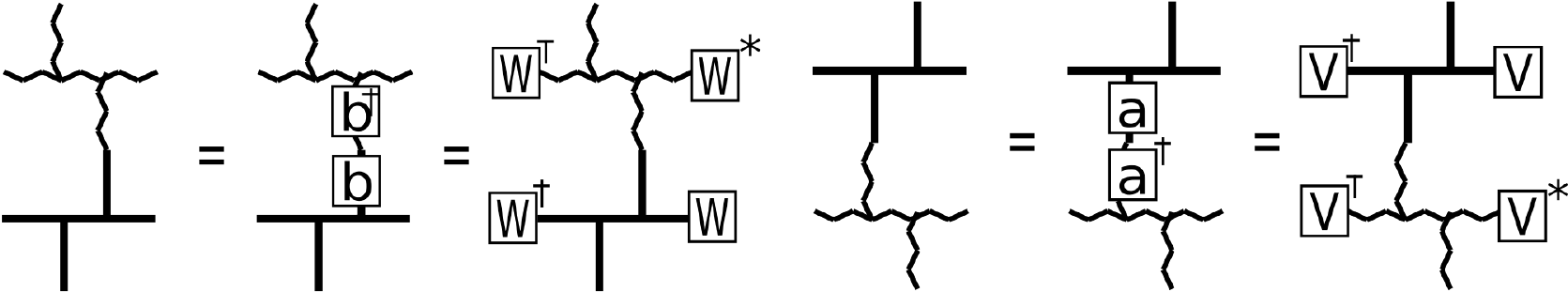}
(a)
 \includegraphics[width=.47\textwidth]{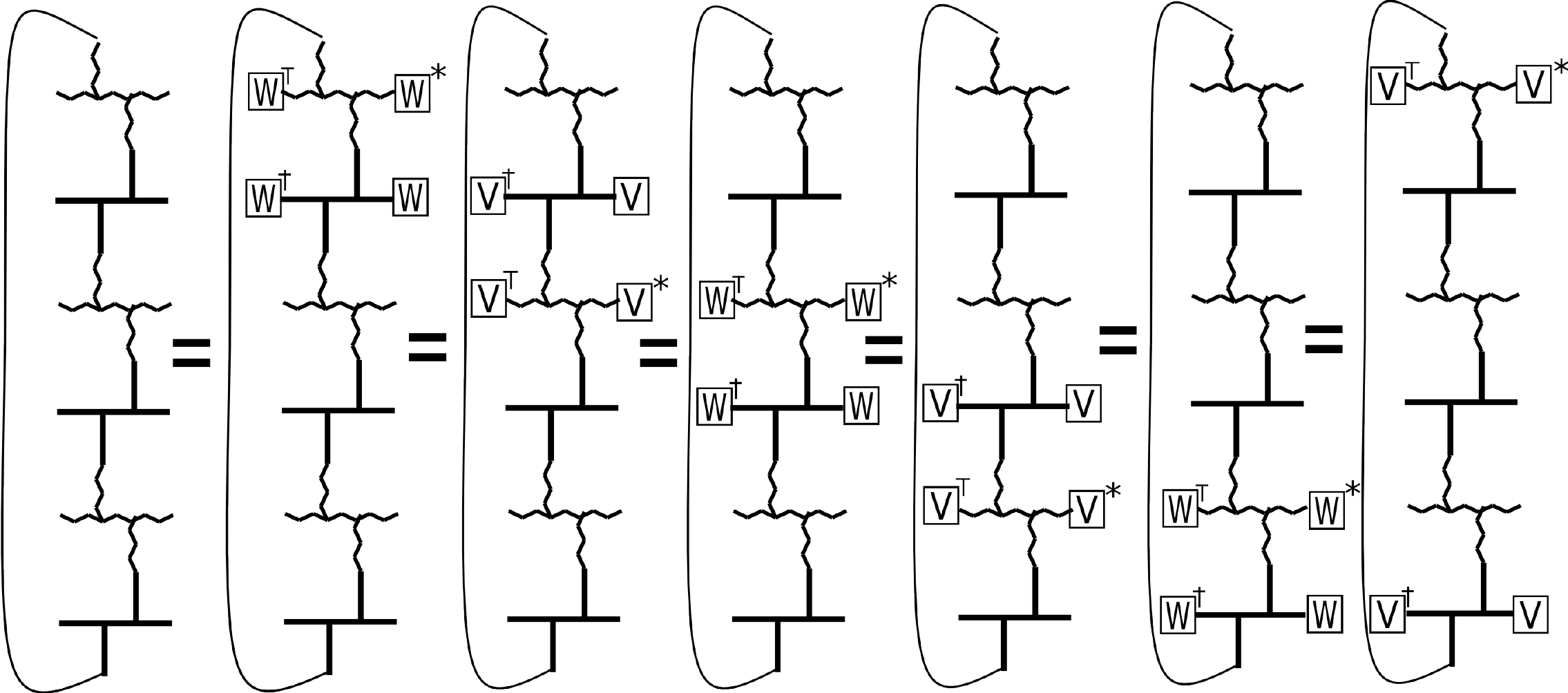}
 (b)
 \end{center}
 \caption{Transfer matrix and its symmetry. The left-most diagram of (b) represents the transfer matrix $\mathbb T_3$ for $\rho^3$.
 The ensuing equalities are direct consequences of the symmetry lifted to the virtual level.
 This implies that the transfer matrix $\mathbb T_\alpha$ has degeneracy $q^{\alpha-1}$ for some integer $q>1$.}
 \label{fig:transfer-symmetry}
\end{figure}

For R\'enyi entropy computations,
we need to evaluate $\Tr(\rho^\alpha)$ for a positive integer $\alpha$.
This amounts to contracting all top vertical bonds with the bottom vertical bonds in Figure~\ref{fig:rhos}.
It is instructive to look at $\rho^2$.
Due to the 1D structure, it is useful to analyze {\em transfer matrix} $\mathbb T$ defined by
the left-most diagram in Figure~\ref{fig:transfer-symmetry}.
For each integer R\'enyi index $\alpha$, there is a transfer matrix $\mathbb T_\alpha$,
which is independent of system size $L$.
Note that the rows and columns of the reduced density matrix $\rho$ are indexed by the physical qudits,
whereas those of the transfer matrix are indexed by the virtual bonds.
We have a trivial yet useful identity:
\begin{equation}
 \Tr(\rho^\alpha) = \Tr( \mathbb T_{\alpha} ^L ).
 \label{eq:trace-is-trace}
\end{equation}
If the eigenvalues of $\mathbb T$ are $\{ \lambda_i \}$,
then $ \Tr( \mathbb T^L ) = \sum_{i=1}^{D^{2\alpha}} \lambda_i^L $,
where $D^{2\alpha}$ is the size of the matrix $\mathbb T$.

There is no guarantee that $\lambda_i$ are all positive;
indeed, in Appendix~\ref{app:intertwinedCluster}
we give an example whose nonzero eigenvalues of $\mathbb T_2$ are
$\frac12,\frac12,\frac12,-\frac12$.
In this case, the entanglement entropy is
\begin{equation}
 S_2(L) = L \log 2 - \begin{cases} \log 4 & \text{if $L$ is even,} \\ \log 2 & \text{otherwise.} \end{cases}
 \label{eq:entropy-twisted-cluster}
\end{equation}

Summarizing, the R\'enyi entropy for $L$ sites is
\begin{equation}
 S_\alpha =  \frac{\log (1/\lambda_1^{(\alpha)}) }{\alpha-1}  L - \frac{\log m}{\alpha-1} + \cdots
 \label{eq:entropy-scaling}
\end{equation}
where $\lambda_1^{(\alpha)}$ is the largest eigenvalue of $\mathbb T_\alpha$,
which is necessarily real positive,
$m=m(L)$ is an integer which may depend on $L$,
and $\cdots$ represents vanishing terms in the large $L$ limit,
The number $m$ is usually the degeneracy of the largest eigenvalue of $\mathbb T$.

\subsection{Degeneracy of the transfer matrix}

We now use the nontrivial SPT to show that the transfer matrix $\mathbb T_\alpha$ has
degeneracy $m \ge q^{\alpha-1}$ for some integer $q> 1$.
The degeneracy bound is uniform to every eigenvalue.
This implies that
\begin{equation}
 \gamma_\alpha = \frac{\log m}{\alpha-1} \ge \log q > 0.
 \label{eq:gammaLowerBound}
\end{equation}

To this end, we rewrite the symmetry lifting Eq.~\eqref{eq:diagSymLifting} as in Figure~\ref{fig:transfer-symmetry}(a),
by which we define the unitaries $V$ and $W$ up to phase factors.
The diagrams in Figure~\ref{fig:transfer-symmetry}(b) follows at once.
The nontrivial SPT implies that
\begin{equation}
 W V = \Omega V W, \quad \Omega = \exp(2\pi i p /q) \neq 1
 \label{eq:Omega}
\end{equation}
where $p$ and $q>1$ are coprime.

Let us define $X_j = W^*_{2j-1} \otimes W_{2j}$
and $Y_j = V_{2j} \otimes V^*_{2j+1}$,
where we have indexed the virtual bonds
from the top to the bottom by integers modulo $2\alpha$.
The index $j$ takes values $1,\ldots, \alpha$.
The symmetries $X_j,Y_j$ of $\mathbb{T}$
form an algebra obeying the following commutation relations.
\begin{align}
 X_j Y_j &= \Omega Y_j X_j \quad (j = 1,\ldots, \alpha) \\
 X_{j+1} Y_j &= \Omega^{-1} Y_j X_{j+1} \quad (X_{\alpha+1} = X_1). \nonumber
\end{align}
All other commutators among $X_i, Y_j$ are vanishing.
They can be rearranged as follows to determine a minimal representation.
Since $X_1 X_2 \cdots X_\alpha$ and $Y_1 Y_2 \cdots Y_\alpha$
are in the center of the algebra,
we take out $X_\alpha$ and $Y_\alpha$ from the generating set,
and do not consider them any more.
\begin{align}
Z_j &:= X_1 X_2 \cdots X_j \quad \text{ for } j = 1,\ldots, \alpha -1 \nonumber \\
Z_j Y_j &= \Omega Y_j Z_j \\
[Z_j, Y_{j'}] &= 0 \quad \text{ if } 1 \le j \neq j' \le \alpha -1 \nonumber
\end{align}
Since $Z_j,Y_j$ generate the same algebra as $X_j,Y_j$ do,
it is clear that the minimal representation of the symmetry
algebra generated by $X_i, Y_i$ has dimension $q^{\alpha -1}$
where $q$ is the multiplicative order of $\Omega$.
It follows from the nontrivial SPT assumption that $q >1$.
We have proved Eq.~\eqref{eq:gammaLowerBound} for any integer $\alpha > 1$.
Therefore, we have found the promised condition
that if the reduced 1D chain is a nontrivial SPT under a product group,
the cylinder extrapolation method will give a nonzero sub-leading term of entanglement entropy.

Note that we have not used all physical symmetry elements.
Sufficient is only one pair of commuting physical symmetry operators
that are lifted to non-commuting virtual unitaries.
The actual degeneracy may be even larger.

\subsection{Cluster state}

Let us apply our general analysis to the previous example of the cluster state.
It turns out that the 1D cluster state has $\mathbb{Z}_2 \times \mathbb{Z}_2$ symmetry under which it is a nontrivial SPT, and the nonzero $\gamma$ will be a consequence of this.
This example is the simplest possible.
We will need to construct an MPS representation,
and identify the projective symmetry on the virtual level.

To find an MPS representation,
we write the wave function of the 1D cluster state in $\sigma^z$ basis
using Eq.~\eqref{eq:ProdCPhaseU}.
\begin{equation}
\langle{\cdots s_{j-1} s_j s_{j+1} \cdots } | {\psi} \rangle = (-1)^{\sum_j s_j s_{j+1} } / \sqrt{2^{2L}}
\end{equation}
where $s_j = 0,1$ and $2L$ is the number of qubits.
It is therefore sufficient for the local tensor $A_j$ at site $j$ to take value $-1$
if $s_j = s_{j+1} = 1$ and $+1$ otherwise. The following tensor satisfies this condition.
\begin{align}
 \mathbf M = \sum_{s = 0,1} \frac{(-1)^{s s'}}{\sqrt{2}} \ket{ s } \otimes \ket{s} \bra{s'}
 \label{eq:mpsBFM}
\end{align}
The physical qubit at site $j$ is synchronized with the left virtual bond,
and thus $s'$ is the state of the physical qubit at site $j+1$
upon contracting the virtual bonds.
The local tensor $\mathbf M$ correctly describes the cluster state, but the symmetry action will not be on-site.
So, we block two neighboring physical qubits as one super-site
and write the MPS representation as
\begin{align}
\mathbb M = \sum_{s_L, s_R = 0,1; s'=0,1} \frac{(-1)^{s_L s_R + s_R s'}}{2} \ket{s_L s_R} \otimes \ket{s_L} \bra{s'}
\label{eq:mpsBoldM}
\end{align}
It is easier to determine the symmetry once we rewrite $\mathbb M$ as a collection of matrices.
\begin{align}
\mathbb M^{(++)} &= \frac12 \begin{pmatrix} 1 & 0 \\ 0 & 1 \end{pmatrix}, &
\mathbb M^{(+-)} &= \frac12 \begin{pmatrix} 0 & 1 \\ 1 & 0 \end{pmatrix}, \\
\mathbb M^{(-+)} &= \frac12 \begin{pmatrix} 1 & 0 \\ 0 & -1 \end{pmatrix}, &
\mathbb M^{(--)} &= \frac12 \begin{pmatrix} 0 & 1 \\ -1 & 0 \end{pmatrix}, \nonumber
\end{align}
where $\ket \pm = (\ket 0 \pm \ket 1) / \sqrt{2}$.
Notice if there was no $\mb{M}^{(++)}$,
the remaining three tensors form an MPS description of the AKLT state.\cite{AKLT1988CMP}
The action of $\sigma^x$'s on every other physical qubit forms a group $G = \mathbb Z_2 \times \mathbb Z_2$, where
the first component $\mathbb Z_2$ is implemented as
\begin{equation}
\mathbb M^{(+\star)} \leftrightarrow \mathbb M^{(+\star)},
\quad \mathbb M^{(- \star)} \leftrightarrow - \mathbb M^{(-\star)},
\end{equation}
and the second $\mathbb Z_2$ is implemented as
\begin{equation}
\mathbb M^{(\star +)} \leftrightarrow \mathbb M^{(\star+)},
\quad \mathbb M^{(\star -)} \leftrightarrow - \mathbb M^{(\star -)}.
\end{equation}
These transformations can be enacted by conjugations
by $\sigma^x$ and $\sigma^z$.
The conjugations on the virtual level does not change the state at all,
and therefore $G$ is a symmetry of the cluster state.

It is evident now that the symmetry we just identified is in accordance with Eq.~\eqref{eq:symmetryLifting}
with $\eta_g$ being trivial. The commuting symmetry action $G$ on the physical level
is lifted to a noncommuting symmetry $D_4 = \langle \sigma^x, \sigma^z \rangle$ on the virtual level. In fact, it is known $H^2(\mb{Z}_2\times\mb{Z}_2;U(1))=\mb{Z}_2$,
and this representation is precisely a projective representation of the group $\mb{Z}_2\times\mb{Z}_2$
and the cluster state is a nontrivial SPT associated with it.
The phase factor of Eq.~\eqref{eq:Omega} is $\Omega = -1$,
and the result of the previous section implies that $\gamma \ge \log 2$.
Indeed, we have $\gamma = \log 2$ in Eq.~\eqref{eq:ZXZ-entropy}.

\section{Generic behavior of $\gamma$: replica correlation length}
\label{sec:RCL}

In the previous section we attributed
the subleading term $\gamma$ of the entanglement entropy
to the degeneracy of the transfer matrix for R\'enyi entropies.
Generically when there is no degeneracy, the entanglement entropy would be
\begin{equation}
S_\alpha = \frac{ | \log \lambda_1 | }{\alpha-1} L + O( r^L ),
\quad r = \left| \frac{\lambda_2}{ \lambda_1} \right| = e^{-1/\xi_\alpha} < 1.
\label{eq:lengthScaleXiE}
\end{equation}
where $\lambda_2$ is the second largest eigenvalue of the transfer matrix. The subleading term converges to zero with an characteristic length scale $\xi_\alpha$.
This indicates that without a symmetry protection the subleading term $\gamma$ should be zero
for long chains.
However, it is important to remark that $\xi_\alpha$ has little to do with
the correlation length $\xi$.
Indeed, being the ground state of a commuting Hamiltonian, the cluster state has the correlation length $\xi = 0$.
Nonetheless, there is degeneracy in the transfer matrix for R\'enyi entropy,
which means that $\xi_\alpha = \infty$ for any $\alpha = 2,3,4,\ldots$.

To understand the generic finite size effect more closely,
we consider a deformed cluster state $\ket{ \theta }$ specified by an
angle $\theta \in [0, \pi]$.
The deformation is achieved by replacing the $-1$ with $e^{i \theta}$
in the MPS representation Eq.~\eqref{eq:mpsBoldM}.
This amounts to transforming the state $\bigotimes_i \ket +$ by
a two-qubit unitary $U(\theta) = \mathrm{diag}(1,1,1,e^{i \theta})$
instead of $U$ in Eq.~\eqref{eq:defnCPHASE}.
It is anyway a transformation by a small-depth quantum circuit from a state of no correlation,
the resulting state $\ket \theta$ has correlation length identically zero.
Clearly, $\theta = \pi$ reproduces the previous cluster state.

The tensor network in Figure~\ref{fig:rhos} for $\ket \theta$ can be explicitly evaluated,
although the computation becomes more complicated as $\alpha$ is increased.
The eigenvalues $\lambda_{1,2,\ldots}$
of the transfer matrix $\mathbb T_2$ for $\rho^2$ can be easily
computed by a computer algebra system.
The result is that
\begin{align}
\lambda_1 &= \frac{3+y^2+(y+1) \sqrt{(y-1)^2+4}}{8} \nonumber \\
\lambda_2 &= \frac{3+y^2-(y+1) \sqrt{(y-1)^2+4}}{8}  \label{eq:ThetaStateTEigenvalues}\\
\lambda_3 &= \frac{y^2-1}{4} , \quad \text{where } y = \cos \theta \nonumber
\end{align}
and all other eigenvalues vanish for any $\theta$.
The largest eigenvalue $\lambda_1$ is non-degenerate unless $\theta = \pi$,
at which the symmetry $G = \mathbb Z_2 \times \mathbb Z_2$ is restored.
The $\theta = 0$ point is also $G$-symmetric;
however, the symmetry is lifted to an abelian virtual symmetry,
and hence the state is a trivial SPT.
The ratio of the second largest eigenvalue to $\lambda_1$ can be any value between $0$ and $1$.
In other words, the length scale $\xi_{\alpha=2}(\theta)$ interpolates from $0$ to $\infty$ continuously,
while the correlation length $\xi$ is held at zero.

Kitaev and Preskill~\cite{KitaevPreskill2006Topological} gave an argument
that the subleading term $\gamma$ can be robustly defined by taking a linear combination of entanglement entropies.
There, it was essentially used that a small change in a region $A$ far from a region $B$
leaves the following combination invariant:
\[
 \Delta S(A) - \Delta S(A \cup B) \simeq 0 .
\]
This is false when $A \cup B$ happen to include exactly a half of a nontrivial 1D SPT chain as in Bravyi's example
in Sec.~\ref{sec:bravyi}.
From our consideration, generically,
the distance between the region $B$ and the region at which the change occurs
should be compared to the replica length scale $\xi_\alpha$ of Eq.~\eqref{eq:lengthScaleXiE},
not to the usual correlation length.
It should be made clear that we did not prove that
a small $\xi_\alpha$ implies that the cylinder extrapolation method or Kitaev-Preskill prescription
gives the total quantum dimension of the topological particle content.
Rather, we showed a short correlation length does not imply
that TEE that results from the cylinder extrapolation method gives the total quantum dimension.
We gave an evidence for the conjecture
that a short replica correlation length
would imply the validity of the cylinder extrapolation method.

\section{Replica correlation functions}
\label{sec:RCF}

\begin{figure}
\centering
\includegraphics[width=.47\textwidth]{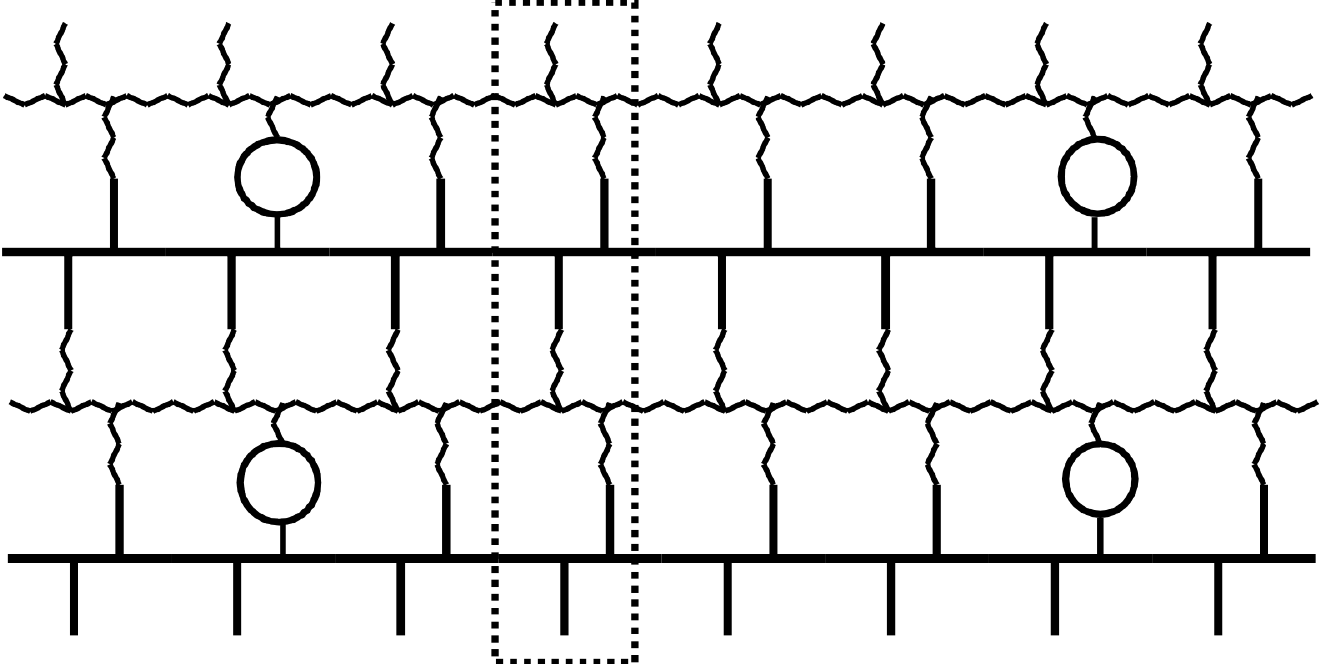}
\caption{Measuring replica correlation functions.
Observables are inserted in the circles.
Even if the global state $\rho_{AB}$ has a short correlation length,
the positive semi-definite operator $\rho_A^2$, treated as a normalized state $\rho_A^2 / \Tr(\rho_A^2)$,
may have much longer correlation length. The latter length scale, which we call as the replica correlation length,
can be simply measured in numerical calculations, and is the relevant length scale for the subleading term
in the entanglement entropy.}
\label{fig:measuringXi2}
\end{figure}

We can actually probe the replica correlation length $\xi_\alpha$
by a {\em replica correlation function}
on the original 2D state, without reducing it into a 1D chain.
Let $\psi = \ket \psi \bra \psi$ denote the density matrix of a state.
Given a two copies of a state $\psi^{\otimes 2}$,
and a bipartition $A \sqcup B$ of the system,
we are formally provided with four subsystems $A_1, B_1, A_2, B_2$.
Define $\mathcal F_A$ to be the swap operator between $A_1$ and $A_2$.
If $\psi_B = \Tr_A[ \psi ]$ is the reduced density matrix for $B$,
it holds that
\begin{align}
\Tr[ \psi^{\otimes 2} \mathcal F_A ( O_{B_1} \otimes O_{B_2} ) ]
&= \Tr[ \psi_B O_{B_1} \psi_B O_{B_2} ] \nonumber \\
&=: \Tr[\psi_B^2] \langle \mathbf O_B \rangle_{\alpha=2}
\label{eq:replica-exp}
\end{align}
for an arbitrary observable $\mathbf O_B = O_{B_1} \otimes O_{B_2}$ on the subsystem $B$.
By Eq.~\eqref{eq:replica-exp}, we have defined $\langle \mathbf O_B \rangle_{\alpha=2}$.
For an observable $\mathbf O_B = O_{B_1} \otimes O_{B_2}$,
we define a {\em replica} correlation function
\begin{align} \label{eq:replicacorrelationfn}
 &\Cor_{\alpha=2}(\mathbf O_B )(x)\\
 &:= \langle \mathbf O_B(i=0) \mathbf O_B(i=L) \rangle_{\alpha=2} \nonumber \\
 &\quad \quad - \langle \mathbf O_B(i=0) \rangle_{\alpha=2} \langle \mathbf O_B(i=L) \rangle_{\alpha=2} . \nonumber
\end{align}
The slowest possible decay of $\Cor_{\alpha=2}$ is determined by
the length scale $\xi_2$ of Eq.~\eqref{eq:lengthScaleXiE},
and this is actually achievable.

We prove this claim by an example.
We will show the cluster state $\ket{\theta = \pi}$
has a nonzero constant replica correlation function that does not decay at all
as a function of the distance between observables.

In addition, we will calculate the replica correlation functions
for the $\mathbb Z_N$ gauge theory ground state in two dimensions,
and the double semion ground state also in two dimensions.
We find that the replica correlation function decays abruptly;
the replica correlation length is zero.
The purpose of this computation is to show that
genuinely topologically ordered phases
that give nonzero $\gamma$
do have representative wave functions with fast decaying replica correlation functions.
We expect that every Levin-Wen wave function\cite{LevinWen2005String-net}
would have a small replica correlation length.
This strongly suggests that the replica correlation function
{\em can} be used to determine when one should rely on the $\gamma$ value
obtained from, e.g., DMRG computation.

\begin{figure}[tb]
\centering
 \includegraphics[width=.45\textwidth]{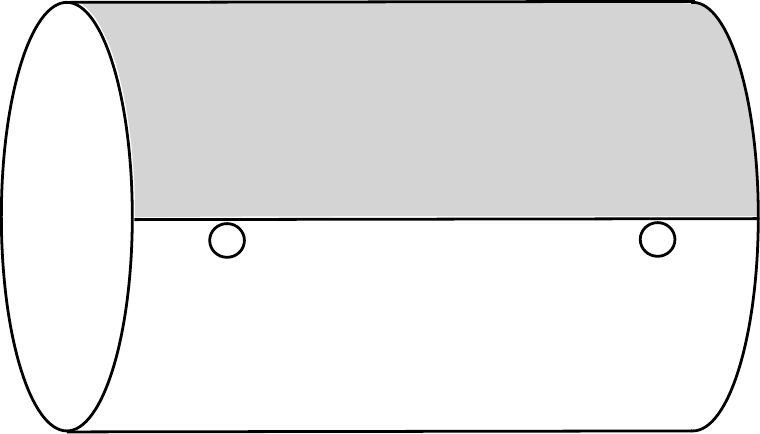}
\caption{Replica correlation function calculation.
 One prepares two copies of the state, and apply the swap operator on the shaded region;
 insert observables in the circles, and compute the overlap with the original, unswapped
 state. The overlap is generally exponentially small in the boundary length of the shaded region,
 but after normalization this reveals the replica correlation length.
}
 \label{fig:ReplicaCorDecay}
\end{figure}

In a DMRG calculation for a (isotropic) 2D state,
we propose a measurement of $\xi_{\alpha=2}$ by the following steps.
One prepares two copies of the state, and inserts
the swap operator on the half strip of the state.
Then, one measures the correlation function for a pair of point observables
inserted near the region where swap operator is applied.
See Fig.~\ref{fig:ReplicaCorDecay}.

\subsection{Flat entanglement spectrum}
\label{sec:FlatES}

Before we delve into the replica correlation function calculations,
we review a technique and expression for special reduced density matrices
that are proportional to projectors.\cite{
KlappeneckerRoetteler2002stabilizer,
VandenNestDehaeneDeMoor2004Local}
This includes the cluster state and the ground states of $\mathbb Z_N$
gauge theory that are eigenstates of string operators.
The technique here will be used crucially in later calculations.
We follow Proposition 9 and Corollary 10 of Ref.~\onlinecite{LindenMatusRuskaiEtAl2013Quantum}.

Define $N \times N$ matrices as
\begin{align}
X_N = \sum_{j \in \mathbb Z_N} \ket{j+1}\bra{j}, \quad
Z_N = \sum_{j \in \mathbb Z_N} e^{2\pi i j/N} \ket j \bra j .
\label{eq:XZmatrices}
\end{align}
Any product $P = X_N^n Z_N^m$ of these matrices has a property that
\begin{align}
\Tr P = \begin{cases} N & \text{if }n=m=0 \in \mathbb Z_N, \\
 0 &\text{otherwise.} \end{cases}
\label{eq:PauliTrace}
\end{align}
Now, consider any multiplicative group $\cG$ generated
by tensor products of these matrices together with the phase factor $e^{2\pi i /N}$
on $(\mathbb C^N)^{\otimes n}$.
Examples are the multiplicative group generated by the term of the Hamiltonian of
the cluster state, and of the $\mathbb Z_N$ gauge theory as
in \eqref{eq:ZNHamiltonian} below.
If $\cG$ is abelian, there exists a common eigenstate
$\ket \psi \in (\mathbb C^N)^{\otimes n}$
of all elements of $\cG$.

Suppose that
\begin{itemize}
 \item there is a unique eigenstate $\ket \psi$ of eigenvalue $+1$ for all $g \in \cG$.
\end{itemize}
Then, $\cG$ cannot contain any pure scalar element $\eta \neq 1$
because such a scalar can only have eigenvalue that is not $1$.
The projector onto $\ket \psi$ can be written as
\begin{align}
\ket \psi \bra \psi = \Pi_{\cG} := \frac{1}{|\cG|} \sum_{g \in \cG} g .
\end{align}
Because of \eqref{eq:PauliTrace},
any non-identity element of $\cG$ has zero trace.
Hence, taking the trace on both sides we see
\begin{align}
1 = \frac{1}{|\cG|} \Tr(I) = \frac{N^n}{|\cG|};
\end{align}
that is, the order of the group must be
the full dimension of the Hilbert space $(\mathbb C^N)^{\otimes n}$.

Now, divide the system into two subsystems $A$ and $B = A^c$
so that $(\mathbb C^N)^{\otimes n} = (\mathbb C^N)^{\otimes |A|} \otimes (\mathbb C^N)^{\otimes |B|}$.
Tracing out $A$ from the density matrix $\ket \psi \bra \psi$,
we obtain the reduced density matrix for $B$:
\begin{align}
 \rho_B = \Tr_A \Pi_{\cG} = \frac{1}{N^{|A|+|B|}} \sum_{g \in \cG} \Tr_A (g)
\end{align}
Again due to \eqref{eq:PauliTrace},
$\Tr_A(g)$ is zero unless $g$ acts on $A$ by the identity (supported on $B$),
in which case $\Tr_A(g) = (g|_B) N^{|A|}$.
The elements of $\cG$ that are supported on $B$ form a subgroup $\cG_B$,
and we can write
\begin{align}
 \rho_B
 = \frac{N^{|A|}}{N^{|A| + |B|}} \sum_{g \in \cG_B} g
 = \frac{|G_B|}{N^{|B|}} \underbrace{\frac{1}{|G_B|} \sum_{g \in \cG_B} g}_{\Pi_{\cG_B}}.
\end{align}
This implies that $\rho_B$ is proportional to the projector $\Pi_{\cG_B}$
of rank $N^{|B|} / |\cG_B|$.
In other words, the entanglement spectrum (the eigenvalues of $\rho_B$) is flat
and the entanglement entropy (von Neumann or R\'enyi) is
\begin{align}
 S(B) = |B| \log N - \log |\cG_B|.
\end{align}
This expression for the entropy is also derived in Ref.~\onlinecite{FattalCubittYamamotoEtAl2004Entanglement,HammaIonicioiuZanardi2005}.

\subsection{Cluster state}

In this subsection we calculate the replica correlation length
of the 2D cluster state and its deformed cousins,
by considering the behavior of the replica correlation functions.
We will find observables that achieve the slowest possible decay of replica correlation functions.
Since a local observable is mapped to a local observable
under finite depth quantum circuits (local unitaries),
we may study the replica correlation function
after simplifying the state by local unitaries.
This means that we can focus on the 1D cluster state
that is reduced from the 2D cluster state.

Consider the cluster state $\ket{\theta=\pi}$ on $2L$ spins
with the periodic boundary condition.
As before, let $B$ be the region that contains every other spin,
in total of $L$ spins.
The reduced density matrix $\psi_B$ has a flat eigenvalue spectrum;
$\psi_B$ is proportional to a projector.
\begin{align}
 \psi_B = \frac{1}{2^L} \left( I + \prod_{i \in B} \sigma^x_i \right)
 \label{eq:rho-cluster}
\end{align}
where the number of nonzero eigenvalues is $M = 2^{L-1}$.
See the previous subsection~\ref{sec:FlatES} for a derivation.

Let $\mathbf O_B(i) = \sigma_i^z \otimes \sigma_i^z$ be an observable
for two copies of the state.
The normalization factor $\Tr[ \psi_B^2 ]$ in Eq.~\eqref{eq:replica-exp}
is equal to $1/M$.
Thus,
\begin{align}
&\langle \mathbf O_B(0) \mathbf O_B(i) \rangle_{\alpha=2} \nonumber\\
&= M \Tr[ \psi_B \sigma_0^z \sigma_i^z \psi_B \sigma_0^z \sigma_i^z]\nonumber\\
&= M \Tr[ \psi_B \sigma_0^z \sigma_i^z  \sigma_0^z \sigma_i^z \psi_B] \\
&= M \Tr[ \psi_B \psi_B] \nonumber\\
& = 1 \nonumber
\end{align}
where the second equality is because $\sigma^z_0 \sigma^z_i$ commutes with $\psi_B$.
On the other hand,
\begin{align}
&\langle \mathbf O_B(j) \rangle_{\alpha=2}\nonumber\\
&= M \Tr[ \psi_B \sigma^z_j \psi_B \sigma^z_j ]\\
&= 0\nonumber
\end{align}
because $\psi_B \sigma^z_j \psi_B = 0$.
Therefore, the replica correlation function reads
\begin{align}
\Cor_{\alpha=2}(\mathbf O_B(0), \mathbf O_B(x) ) = 1
\end{align}
independent of the separation $x$.

For generic values of $\theta$,
we numerically checked that replica correlation functions of generic observables
on the state $\ket \theta$ decay according to the finite replica correlation length
calculated from \eqref{eq:ThetaStateTEigenvalues}.
We emphasize once again that the usual correlation length $\xi$ measured by
\[
\bra \theta K_1 K_2 \ket \theta - \bra \theta K_1 \ket \theta \bra \theta K_2 \ket \theta
\]
is identically zero for any observables $K_1$ and $K_2$
for the state $\ket \theta$ for any $\theta$.

\subsection{$\mathbb Z_N$ gauge theory} \label{sec:ZN-xi2}

The replica correlation length/function
is introduced to pick up a fine detail of a state,
and therefore we calculate it for a particular ground state
of an exactly soluble model of the $\mathbb Z_N$ gauge theory in the deconfined phase.
Unlike the 2D cluster state,
abelian discrete gauge theory ground states,
as well as the double semion state of the next subsection,
requires deep quantum circuit to disentangle,%
\cite{BravyiHastingsVerstraete2006generation,Hastings2010Locality,Haah2014invariant}
so we are forced to work with the 2D state directly in the replica correlation functions.
The lattice of the model is not too important,
but we consider the square lattice in two dimensions.
The Hamiltonian is sum of star terms (gauge transformation), and plaquette terms (flux):
\begin{align}
 H_{\mathbb Z_N} =
 &- \sum_s X_{s,\text{east}} X_{s,\text{north}}
  X_{s,\text{west}}^\dagger X_{s,\text{south}}^\dagger \nonumber \\
 &- \sum_p Z_{p,\text{south}} Z_{p,\text{east}}
  Z_{p,\text{north}}^\dagger Z_{p,\text{west}}^\dagger.
\label{eq:ZNHamiltonian}
\end{align}
where $s$ denotes a site (vertex) and $p$ denotes a plaquette (face),
and $X = X_N$, $Z = Z_N$ of \eqref{eq:XZmatrices}.
The ground state is an equal amplitude superposition
of ``loop'' configurations,
where the loops come in $N$ types
and obey the group law of $\mathbb Z_N$.

When put on a thin torus as in Fig.~\ref{fig:ReplicaCorDecay},
the Hamiltonian $H_{\mathbb Z_N}$
has an $N^2$-fold degenerate ground space.
As the DMRG algorithm is biased to states with minimal entanglement
across the circumferential cut,\cite{JiangWangBalents2012}
we consider the state $\ket \psi$ that has $+1$ eigenvalue
of the $Z$-type string operator
and $+1$ of $X$-type string operator along the circumference
(the shortest nontrivial loop).

For this state $\ket \psi$,
which is the unique common $(+1)$-eigenvector of a commuting set of
tensor products of $X$ and $Z$ matrices,
the entanglement spectrum
for any bipartition is flat,
and the reduced density matrix $\psi_B$
is proportional to a projector
\begin{align}
 \Pi = \frac{1}{|\cG_B|} \sum_{g \in \cG_B} g.
\end{align}
Here, the group $\cG_B$
is a subgroup of the multiplicative group $\cG$ that is generated by all Hamiltonian terms.\footnote{
For other ground states that are eigenstates of string operators
on the orthogonal loop, this is not the case
and one has to include the topologically nontrivial string operators in $\cG$.
}
$\cG_B$ consists of all elements of $\cG$ that is supported on $B$.
(See the previous subsection~\ref{sec:FlatES}.)

In fact, $\cG_B$ is generated precisely by the Hamiltonian terms supported on $B$.
To see this, suppose $g \in \cG_B$.
Since $\cG$ is abelian,
the operator $g$ can be written as a product of closed $Z$-loop operators
and closed $X$-loop operators.
These loop operators are contractible and supported on $B$,
so each $X$- or $Z$-loop can be deformed to vanish
by multiplying the smallest loop operators,
which is exactly the Hamiltonian terms on $B$.
This implies that $g$ is a product of the Hamiltonian terms on $B$.
$\cG_B$ having a local generating set
is an important difference from the cluster state,
for which the density matrix \eqref{eq:rho-cluster}
is the sum over a group with \emph{no local} generators.

Note that since $\Pi$ is a sum of all elements of a group, we see
\begin{align}
 \Pi g = \Pi = g \Pi
 \label{eq:absorbG}
\end{align}
for any element $g \in \cG_B$.
We claim that for any operator $O$ there exists $\tilde O$
of the {\em same} support such that
\begin{align}
 \Pi O \Pi = \Pi \tilde O \Pi, \quad \text{ and } \quad [\Pi, \tilde O] = 0.
 \label{eq:ActionCommCondition}
\end{align}
To construct $\tilde O$,
let $\cG_i \le \cG_B$ be the group generated by the Hamiltonian terms
that overlap the site (or region) $i$ on which $O$ is supported.
Let
\begin{align}
\tilde O = \frac{1}{|\cG_i|} \sum_{g \in \cG_i} g O g^{-1}.
\label{eq:symmetrization}
\end{align}
The support of $\tilde O$ is the same as that of $O$
because every element $g \in \cG_i$ is a tensor product unitary operator.
Any tensor component of $g$ that acts outside of the support of $O$
cancels in the expression $g O g^{-1}$.
Hence, every summand $g O g^{-1}$ has the same support as $O$,
so is the sum $\tilde O$.
Note that $\tilde O$ is zero if, e.g., $O$ anti-commutes with some $g \in \cG_i$.

Next, we verify \eqref{eq:ActionCommCondition}.
\begin{align}
\Pi \tilde O \Pi
&=
\frac{1}{|\cG_i|} \sum_{g \in \cG_i} \Pi g O g^{-1}\Pi  \nonumber\\
&= \frac{1}{|\cG_i|} \sum_{g \in \cG_i} \Pi O \Pi \label{eq:SameActionProof}\\
&=\Pi O \Pi, \nonumber
\end{align}
where we used \eqref{eq:absorbG}.
In addition, $g \tilde O g^{-1} = \tilde O$ if a Hamiltonian term $g \in \cG_i$ overlaps $i$
by \eqref{eq:symmetrization}.
If $h \in \cG_B \setminus \cG_i$ is a Hamiltonian term that does not overlap $i$,
then $h$ commutes with $g \in \cG_i$ because $\cG_B$ is abelian,
and $h$ also commutes with $O$ trivially,
so $h$ commutes with $\tilde O$.
Since $\cG_B$ is generated by Hamiltonian terms,
$\tilde O$ commutes with every element of $\cG_B$.\footnote{
Even if a topologically nontrivial string operators are in $\cG_B$,
this is still true
as the long string generator $s$ of $\cG$ can always chosen to be
commuting with $O$ so that $s O s^{-1} = O$.
Note that a Hamiltonian term that overlaps $i$ but is not in $\cG_B$
may \emph{not} commute with $\tilde O$.
}
Since the projector $\Pi$ is the sum of all group elements of $\cG_B$,
it follows that
\begin{align}
[\Pi, \tilde O] = 0.
\end{align}
Moreover, if $O$ and $O'$ are far separated so that
no generator of $\cG_i$ or $\cG_{i'}$ overlaps with both $O$ and $O'$,
then
\begin{align}
\tilde O \tilde O'
&= \frac{1}{|\cG_i| \cdot |\cG_{i'}|} \sum_{g \in \cG_i} \sum_{h \in \cG_{i'}} (gO g^{-1})( h O' h^{-1})\nonumber\\
&= \frac{1}{|\cG_i|\cdot |\cG_{i'}|} \sum_{g \in \cG_i} \sum_{h \in \cG_{i'}} gh O  O' (gh)^{-1}\label{eq:TildeSeparation}\\
&= \frac{1}{|\cG_i \times \cG_{i'}|} \sum_{g \in \cG_i \times \cG_{i'}} g O  O' g^{-1}\nonumber\\
&= \widetilde{OO'}.\nonumber
\end{align}

Consider arbitrary observables $\mathbf O = \sum_a P_a \otimes Q_a$ near a site $i$
and $\mathbf O' = \sum_b P'_b \otimes Q'_b$ near a site $i'$ on the two copies of the state.
Suppose $i$ and $i'$ are sufficiently separated, say by 5 lattice spacing, so that
no term in the Hamiltonian $H_{\mathbb Z_N}$
overlaps simultaneously with $\mathbf O$ and $\mathbf O'$.
The normalization factor of \eqref{eq:replica-exp}
is given by $M = \Tr[\psi_B^2]^{-1} = \Tr[ \Pi ] = |\cG_B|$.
Using tilde operators, the replica correlation function
becomes a usual correlation function:
\begin{align}
 &\langle \mathbf{OO'} \rangle_{2} -\langle \mathbf{O} \rangle_2 \langle \mathbf O' \rangle_2 \nonumber\\
 &=  |\cG_B|^{-1} \sum_{a,b} \Tr[ \Pi P_a P'_b \Pi Q_a Q'_b ] \nonumber\\
    &\quad\quad - |\cG_B|^{-2} \sum_{a,b} \Tr[ \Pi P_a \Pi Q_a] \Tr[\Pi P'_b \Pi Q'_b]\nonumber\\
 &=  |\cG_B|^{-1} \sum_{a,b} \Tr[ \Pi \tilde P_a \tilde P'_b \tilde Q_a \tilde Q'_b ]
 \label{eq:RCFtoCF}\\
  & \quad \quad - |\cG_B|^{-2} \sum_{a,b}\Tr[ \Pi \tilde P_a \tilde Q_a] \Tr[\Pi \tilde P'_b \tilde Q'_b]\nonumber\\
 &= \Cor\left(\sum_a \tilde P_a \tilde Q_a, \sum_b \tilde P'_b \tilde Q'_b \right)\nonumber\\
 &= 0 \nonumber
\end{align}
where in the second equality $\widetilde{P_a P'_b} = \tilde P_a \tilde P'_b$
is because they act on separated spins,
and the last equality is because $H_{\mathbb Z_N}$ is commuting
with locally indistinguishable ground space.

The replica correlation function is not identically zero
when the site $i$ and $i'$ are close.
This is simple.
The replica correlation function becomes the usual correlation function
when $\mathbf O = O \otimes I$.
Consider $O = X_{s,\text{east}} X_{s,\text{north}} + h.c.$ and
$O' =  X_{s,\text{west}}^\dagger X_{s,\text{south}}^\dagger + h.c.$
where $s$ is in the interior of $B$.
Then, $\Tr[OO' \psi_B] = 2$, but $\Tr[O \psi_B] = \Tr[O' \psi_B] = 0$,
so $\Cor(O,O') = 2$.

In conclusion,
we have shown that the replica correlation function for $H_{\mathbb Z_N}$
is not identically zero but decays to zero after separation distance 5.
Therefore, the replica correlation length is zero.
This is in contrast to the cluster state calculation
where the replica correlation function is nonzero and does not decay at all.

\subsection{Double semion model}

Using the similar techinque as in the previous subsection,
we will calculate the replica correlation function
for a version of double semion model on the honeycomb lattice,%
\cite{LevinWen2005String-net,KeyserlingkBurnellSimon2013}
and find that it decays abruptly in the same ways as for the $\mathbb Z_N$ gauge theory.

The lattice consists of 2 spin-$\frac12$'s at each edge.
The two spins on an edge is going to be (energetically) ``synchronized''.
The Hamiltonian is
\begin{align} \label{eq:double-semion-hamiltonian}
 H_{DS} = & - \sum_p \underbrace{ -\prod_{e, e' \in p} \sigma^x_e \sigma^x_{e'} \prod_{e \in \partial p} \sqrt{\sigma^z_e}}_{g_p} + h.c.\nonumber \\
 &- \sum_v \underbrace{\prod_{e \in v} \sigma^z_e}_{g_v} - \sum_e \underbrace{\sigma^z_{e} \sigma^z_{e'}}_{g_e}.
\end{align}
The first term $g_p$ has the minus sign, and
is defined for every hexagon $p$ where $e, e' \in p$ means all 12 spins
on the plaquette, and $e \in \partial p$ means the 6 spins on the legs of the plaquette
that are immediate neighbors of the plaquette.
The second term is defined for every trivalent vertex $v$, and $e \in v$ means
the 3 spins that are immediate neighbors of $v$.
The third term is defined for every edge that contains two spins $e$ and $e'$;
this ferromagnatic term ``synchronizes'' the two spins on the edge.
Our Hamiltonian \eqref{eq:double-semion-hamiltonian}
is slightly different from those in Ref.~\onlinecite{LevinWen2005String-net,KeyserlingkBurnellSimon2013},
but splitting the edge spin into two spins has appeared in Ref.~\onlinecite{LevinWen2006Detecting}.
We are considering the split version because of the simplicity of reduced density matrix expression.

\begin{figure}[h!]
  \centering
\includegraphics[width=0.4\textwidth]{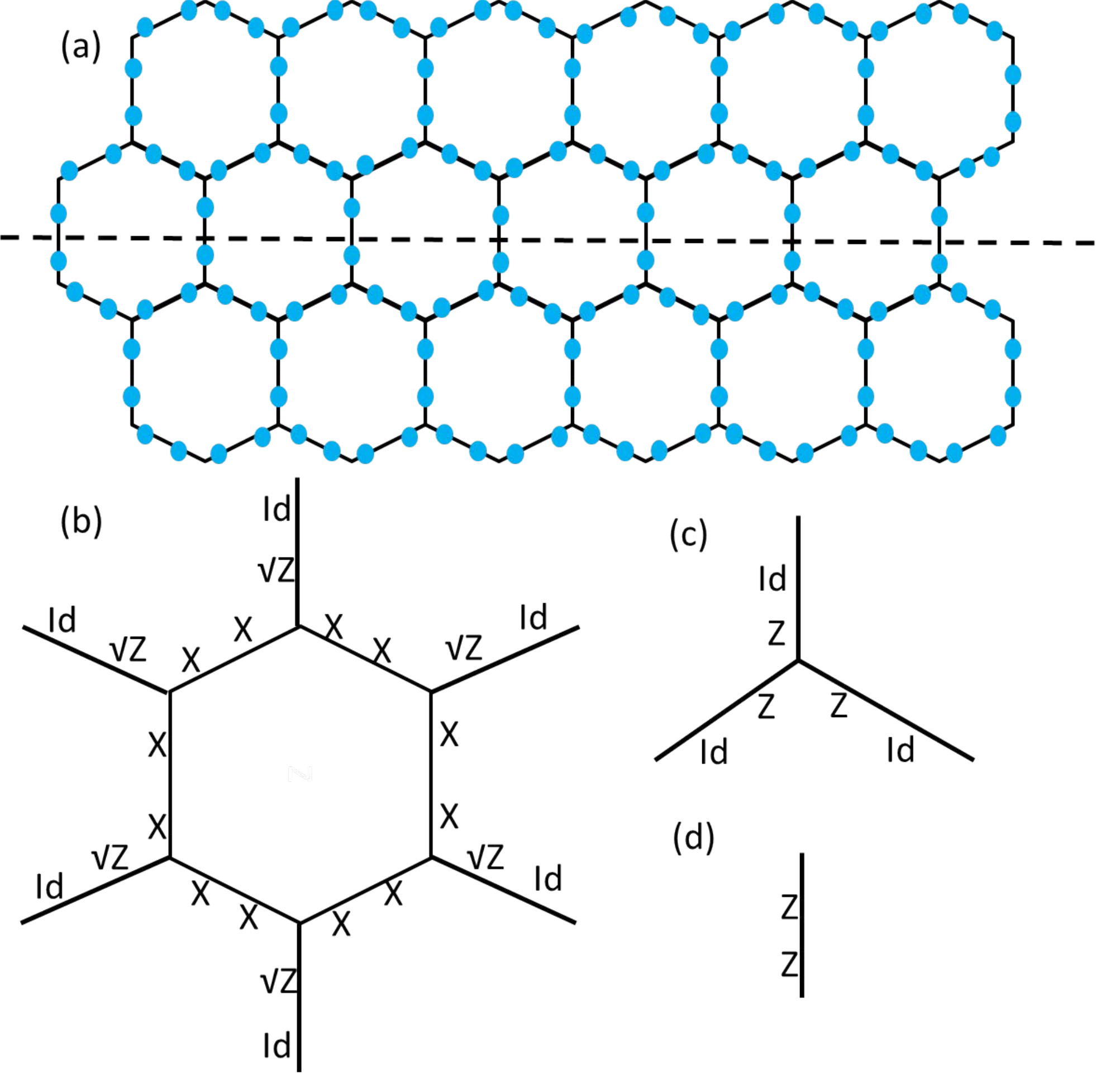}\\
\caption{Double-semion model defined in (\ref{eq:double-semion-hamiltonian}).
(a) shows the lattice configuration and the location of the degrees of freedom.
Each edge of the hexagon accommodates two spins, denoted by the blue dots on the edge.
The dashed line is the entanglement cut.
(b-d) pictorially show the three terms in the Hamiltonian for a plaquette, vertex and edge.
The symbol near a spin denotes the operator acting on this spin,
where $X$ means $\sigma_x$, $Z$ means $\sigma_z$,
$\sqrt{Z}=\mathrm{diag}(1,i)$, and $\mathrm{Id}$ is the identity matrix.
The total Hamiltonian is the summation over all plaquettes,
vertices and edges, with appropriate sign factors defined in
\eqref{eq:double-semion-hamiltonian}.
}
\label{fig:double-semion}
\end{figure}

The plaquette terms $g_p$ commute with the vertex terms $g_v$ and edge terms $g_e$,
but they do not commute among themselves;
however, they do commute on the constrained subspace
where every vertex and edge term takes +1 eigenvalue ($g_v = g_e = 1$).
This is equivalent to the following.
Let $\cG^P$ denote the nonabelian multiplicative
group generated by all $g_p$'s,
and let $\cG^Z$ denote the abelian multiplicative group generated by all $g_v$'s and $g_e$'s.
Also, let $\cG = \cG^P \cG^Z$ denote the group generated by all terms in the Hamiltonian.
Then,
\begin{align}
&g z g^{-1} z^{-1} = 1 \quad \forall g \in \cG^P, z \in \cG^Z \nonumber\\
&ghg^{-1}h^{-1} \in \cG^Z\quad \forall g,h \in \cG^P.
\end{align}
(See also Ref.~\onlinecite{NiBuerschaperNest2015}.)

The ground space of $H_{DS}$ on a torus is four-fold degenerate,
and the minimally entangled states are eigenstates of string operators along
a shortest topologically nontrivial loop.
The string operators whose end points, if open, corresponds to semions
are not too simple (Eq.~9 of Ref.~\onlinecite{KeyserlingkBurnellSimon2013}),
but satisfies an important property that the partial trace is zero.
Let us fix the entanglement cut that passes an array of plaquettes
along a straight line that is orthogonal to an edge.
This entanglement cut passes in between the two spins on the intersecting edges,
which is considered in Ref.~\onlinecite{LevinWen2006Detecting}.
The purpose of this special cut is to have
\begin{align}
 \Tr_A[ g ] = 0 \text{ if $g \in \cG^P$ overlaps with $A$}
\end{align}
since any such $g$ has a $\sigma^x$ or $\sigma^z$ tensor component in addition to $\sqrt{\sigma^z}$
within $A$.
Then, it follows by the same reasoning as in Sec.~\ref{sec:FlatES} that
\begin{align}
 \psi_B = \Tr_A[ \psi ] \propto \sum_{g \in \cG_B} g
\end{align}
where $\cG_{B} \le \cG = \cG^P\cG^Z$,
consists of the elements supported on $B$.\footnote{
The string operator that wraps around the shortest topologically nontrivial loop
of the torus does not enter, because its partial trace is zero.
}
The entanglement spectrum is flat.

The subgroup $\cG_{B}$ is generated by the local Hamiltonian terms supported on $B$.
The reason is similar to that for the previous $\mathbb Z_N$ theory.
Observe that $\sigma^x \sqrt{\sigma^z} \sigma^x = i (\sqrt{\sigma^z})^{\dagger}$.
Hence, any element $g \in \cG = \cG^P\cG^Z$ can be written as
a product of $\prod_i \sigma^x_i$ and $\prod_i \sqrt{\sigma^z_i}^{n_i}$
up to an overall phase factor.
The first factor $\prod_i \sigma^x_i$ has to form a closed loop
since it arises from $g_p$ terms.
The closed loop of $\sigma^x$ must be entirely contained in $B$,
and we can eliminate it by multiplying $g_p$ operators on $B$ to $g$.
Therefore, it suffices for us to show that any ``diagonal'' element
$z \in \cG$ (a product of $\sqrt{\sigma^z}$)
supported on $B$,
is given by a product of $g_v$ and $g_e$ on $B$.
Note that any diagonal element of $\cG^P$ arises from $g_p^2$,
and $g_p^2$ belongs to $\cG^Z$.\cite{NiBuerschaperNest2015}
The group $\cG^Z$ can be viewed as the group of null-homologous $\mathbb Z_2$-loops
on the triangular lattice (the dual lattice of the honeycomb lattice).
Therefore,
if $z \in \cG^Z$ is supported on $B$, then $z$ is a product of $g_v$ and $g_e$ on $B$.
This implies that $g \in \cG$ supported on $B$
is a product of Hamiltonian terms on $B$ up to a phase factor.
The phase factor must be 1 because the group $\cG$
does not contain any nontrivial phase factor.
This completes the reasoning.

As in the previous section,
we can turn any operator $O$ supported on $B$ into $\tilde O$ such that
\begin{align}
\mathrm{support}(O) &= \mathrm{support}(\tilde O ), \nonumber \\
\psi_B O \psi_B &= \psi_B \tilde O \psi_B,\nonumber\\
[\tilde O, \psi_B] &= 0,\label{eq:DSTildeProperties}\\
\widetilde{ O O'} &= \tilde O \tilde O' \quad \text{if separated.}\nonumber
\end{align}
Let $\cG_i^P$ be the group generated by $g_p$ that overlaps $i$
on which $O$ is supported,
and let $\cG_i^Z$ be the group generated by $g_e$ and $g_v$
on spins where $\cG_i^P$ is supported.
The choice of group $\cG_i^Z$ is to have
\begin{align}
 &g z g^{-1} z^{-1} = 1 \quad \forall g \in \cG^P, z \in \cG^Z_i \nonumber\\
 &ghg^{-1} h^{-1} \in \cG^Z_i~~ \forall g \in \cG^P, h \in \cG^P_i .
   \label{eq:effectivelyCommuting}
\end{align}
Now, let $\cG_i = \cG^P_i\cG^Z_i$, and define
$\tilde O$ by the formula \eqref{eq:symmetrization}.
$\tilde O$ has the same support as $O$ is because $\cG_i$
is a group of tensor product unitary operators.
That $\psi_B O \psi_B = \psi_B \tilde O \psi_B$ follows from
a similar equation as \eqref{eq:SameActionProof}
since $\psi_B$ for the double semion state
is also a sum over a group $\cG_B \supset \cG_i$.
As for $[\tilde O, \psi_B] = 0$,
we see $h \tilde O h^{-1} = \tilde O$
if $h \in \cG^P_i$ is one of the plaquette terms $g_P$
that overlap $i$, by definition of $\tilde O$.
If $k \in \cG^P \setminus \cG^P_i$ is a plaquette term that does not overlap $i$,
then
\begin{align}
k \tilde O k^{-1}
&= \sum_{g \in \cG_i}  k g O g^{-1} k^{-1}\nonumber \\
&= \sum_{g \in \cG_i} gk (k^{-1}g^{-1}kg) O (k^{-1}g^{-1}kg)^{-1} (gk)^{-1}\nonumber\\
&= \sum_{g \in \cG_i} g (k^{-1}g^{-1}kg) O (k^{-1}g^{-1}kg)^{-1} g^{-1} \label{eq:DSCommProof}\\
&= \sum_{g \in \cG_i} g O g^{-1}\nonumber\\
&= \tilde O \nonumber
\end{align}
where in the third equality we used \eqref{eq:effectivelyCommuting} and that $[k,O]=0$,
and in the fourth equality we redefined the dummy variable $g$
since $(k^{-1}g^{-1}kg) \in \cG_i$.
If $z$ is one of $g_e$ or $g_v$ terms,
then a similar calculation, simpler than \eqref{eq:DSCommProof},
shows $z \tilde O z^{-1} = \tilde O$.
As we have shown that $\psi_B$ is a sum over a group generated by the terms of $H_{DS}$,
the proof that $[\psi_B,\tilde O]=0$ is complete.
The last property $\widetilde{OO'} = \tilde O \tilde O'$
follows from similar equations as \eqref{eq:TildeSeparation}.

In conclusion,
Eq.~\eqref{eq:RCFtoCF} holds without any modification using \eqref{eq:DSTildeProperties}
for the double semion model under our bipartition.
Therefore, the replica correlation function reduces to a usual correlation function,
and the replica correlation length is zero.

\section{Discussion} \label{sec:discussion}

In this paper we have studied the behavior of the sub-leading term
of the bipartite entanglement entropy of topologically trivial 2D states
calculated by the cylinder extrapolation method,
and found a sufficient condition
under which a topologically trivial state
will give a nonvanishing sub-leading term under this method.
In particular, we showed the bipartite entanglement entropy of a such 2D state
can be reduced to that of a 1D chain under an extensive bipartition.
If this 1D chain is in a nontrivial SPT state under a product group $G=G_1\times G_2$,
where $G_1$ and $G_2$ act exclusively on the two sides of the bipartition,
then a nonvanishing sub-leading term appears in the cylinder extrapolation method.

Our result does not necessarily invalidate the notion of topological entanglement entropy.
In fact, the examples in this paper that are translation-invariant yield
the correct total quantum dimension of the topological phase under the Kitaev-Preskill or Levin-Wen prescription.
Rather, our finding makes it clear that the cylinder extrapolation method
may give a different answer than the Kitaev-Preskill or Levin-Wen prescription in the bulk.

Notice the above condition requires the state be in a nontrivial SPT state of $G$
in the way we described above.
This requirement, where the nontriviality of the state is protected by $G_1$ and $G_2$ simultaneously,
and $G_1$ and $G_2$ act exclusively on the two sides of the bipartition,
is stronger than the general condition of 1D SPT based on group
cohomology.\cite{SchuchPerez-GarciaCirac2010Classifying,Chen2011}
A nontrivial SPT in the general sense can be protected by $G_1$ or $G_2$ alone,
but this is not sufficient to yield a nonvanishing sub-leading term from the cylinder extrapolation method.

We have introduced the replica correlation length/function.
In Sec.~\ref{sec:RCF} we gave an operational meaning to it
and demonstrated that it can be determined numerically.
Though we only discussed $\alpha=2$ replica correlation length/function,
it is straightforward to consider $\alpha >2$ cases
by considering cyclic permutation operators instead of the swap operator.
Our result suggests a conjecture that
the cylinder extrapolation method on minimally entangled states
yields the total quantum dimension of the topological particle content
if the replica correlation length is small compared to the system size.
All our examples of the SPT states under a product group, the $\mathbb Z_N$ lattice gauge theory,
and the double semion model should be read as evidences in favor of this conjecture.

Analogues of TEE for the ground states of gapped Hamiltonians in 3 or higher dimensions
have been proposed.\cite{GroverTurnerVishwanath2011TEE}
Ref.~\onlinecite{GroverTurnerVishwanath2011TEE}
studies various solid torus geometries and identifies multitude of TEEs
that are associated with Betti numbers of the region for which entanglement entropy
is calculated.
Our examples can be generalized to this setting using graph states\cite{HeinEisertBriegel2004graph},
and indeed modifies the subleading constant term
of the entanglement entropy.
If one tries a (hyper-)cylinder extrapolation method
to read off the subleading term,
then our ideas here give translation-invariant states with a modified subleading term.
As remarked before,
our examples in this higher dimensional generalization
will be fine-tuned, but the length scale where
the finite size effect is relevant can be arbitrarily larger
than usual correlation lengths.

Besides, it is natural to consider topological entropy for thermal states.
An immediate problem is that
the entropy of the reduced density matrix of a thermal state obeys a volume law.
This is easily overcome by using mutual information,\cite{CastelnovoChamon2007Entanglement}
which obeys an area law at any nonzero temperature.\cite{WolfVerstraeteHastingsEtAl2008Area}
In 2D, while every known {\em ground} state with a nonzero TEE
requires a local unitary transformation (quantum circuit)
of large depth (linear in system size)
in order to be transformed to a product state,
every Gibbs state at any nonzero temperature of any commuting Hamiltonian
can be transformed by a quantum circuit of small depth (logarithmic in system size)
into a Gibbs state of a classical Hamiltonian.\cite{Hastings2011warmTQO}
This is consistent to the calculation of
topological entropy of the 2D $\mathbb Z_2$
gauge theory at nonzero temperature
where the subleading term of mutual information is shown to
vanish.\cite{CastelnovoChamon2007Entanglement}

In 3D, the entropies on solid torus of $\mathbb Z_2$ lattice gauge theory at nonzero temperature have been calculated.\cite{CastelnovoChamon2008Topological}
It is observed that at low nonzero temperature,
a nonzero subleading term survives
when a certain linear combinations of mutual information is used to cancel extensive parts.
But Ref.~\onlinecite{Hastings2011warmTQO} also shows
that the Gibbs state of this model at nonzero temperature
can be connected to the Gibbs state of a classical Hamiltonian
by a small-depth quantum circuit. (See also Ref.~\onlinecite{SivaYoshida2016Topological}.)
Therefore, on the contrary to the 2D case,
this value being nonzero is not related to topological order
in the sense of generating quantum circuits of large depth.
This means that if we accept the complexity of the generating quantum circuit
as the definition of topological order for thermal states,
then we should conclude that the subleading term of mutual information does not give
an `order parameter' for topological order for thermal states in 3 or higher dimensions.

At nonzero temperature, it appears that our examples
only give a contribution that is exponentially small in the system size to the subleading term
(although the length scale of our effect is still different from the usual correlation length).
The mutual information of the 2D cluster state Eq.~\eqref{eq:clusterHamiltonian}
at finite inverse temperature $\beta$
with respect to the bipartition in Fig.~\ref{fig:triangular-lattice} is
\begin{align}
I(A:B) = 2L (\log 2 + \mathcal O((1-t)\log(1-t))) - \mathcal O(t^{2L})
\end{align}
where $t = \tanh \beta$ is close to but smaller than 1.
The detail of the calculation can be found in Appendix~\ref{app:thermal}.

Finally, we note there exists a notion of localizable entanglement
and associated entanglement length,\cite{Popp2005}
whose divergence is connected to a string order parameter.\cite{Venuti2005}
It is shown that a subclass of our nontrivial 1D SPT can be used as a
perfect quantum repeater.\cite{Else2012}
However, it remains unclear how the entanglement length
is related to our replica correlation length $\xi_\alpha$.
A technical difference is that in our definition $\xi_\alpha$ carries an (artificial) index $\alpha$,
as it is defined by the eigenvalues of the transfer matrix $\mathbb T_\alpha$ for $\rho^\alpha$.

\begin{acknowledgments}
We thank T. Senthil for valuable discussions that have led to this paper,
and S. Bravyi for sharing his unpublished result.
LZ is supported by a US Department of Energy grant de-sc0008739.
JH is supported by Pappalardo Fellowship in Physics at MIT.
\end{acknowledgments}

\appendix

\section{Intertwined chains of cluster state}
\label{app:intertwinedCluster}

We present examples where the sub-leading term $\gamma$ has a system-size-dependent oscillation.
Recall the local tensor for the cluster state Eq.~\eqref{eq:mpsBFM}:
\begin{equation}
 M^{(0)} = \frac{1}{\sqrt{2}} \begin{pmatrix} 1 & 1 \\ 0 & 0 \end{pmatrix}, \quad
 M^{(1)} = \frac{1}{\sqrt{2}} \begin{pmatrix} 0 & 0 \\ 1 & -1\end{pmatrix}
\end{equation}
where the superscripts are the physical indices.
Define a local tensor $K^{(ab)}$ with two physical qubits per site and bond dimension 4
as
\begin{equation}
 K^{(ab)} =  \left( M^{(a)} \otimes M^{(b)} \right)
 \begin{pmatrix}1 & 0 & 0 & 0 \\ 0 & 0 & 1 & 0 \\ 0 & 1 & 0 & 0 \\ 0 & 0 & 0 & 1  \end{pmatrix}  .
 \label{eq:K}
\end{equation}
We trace out the physical qubit $a$ and keep $b$.
One can verify that the transfer matrix for $\alpha = 2$
has four nonzero eigenvalues $\frac12,\frac12,\frac12,-\frac12$.
Thus, the entanglement entropy is
\begin{equation}
S = -\log \left[ 3 \left(\frac12\right)^L + \left( -\frac12 \right)^L \right]
\label{eq:evenodd}
\end{equation}
under the periodic boundary condition.
We omitted R\'enyi index $\alpha$; in fact, this formula is true for any $0 \le \alpha \le \infty$
since the distribution of nonzero Schmidt coefficients (entanglement spectrum) is flat.

This local tensor represents two independent cluster states that are intertwined.
When the chain length $L$ is even, there are two symmetries
supported entirely on $a$ qubits.
But, these symmetries cannot be defined separately when $L$ is odd.
The even-odd behavior may be attributed to the extra symmetry when $L$ is even,
which are not uniformly on-site.

More generally, one can define an MPS such that the subleading term of the entanglement entropy
has periodicity $n$ as a function of chain length $L$ for any positive integer $n$.
Eq.~\eqref{eq:K} will corresponds to $n=2$.
The construction is to imagine a helix of $n$ strands, project it into a plane,
put a qubit to each outer vertex, and interpret the line between vertices as the bonds of the cluster state.
Under periodic boundary condition with length $L$,
the number of distinct strands is given by $\gcd(n,L)$.
Still, the translation-invariance of the state is observed.
The local tensor can be given as
\begin{align}
 K^{(ab)} &=  \left( M^{(a)} \otimes I^{\otimes(n-2)} \otimes M^{(b)} \right)  \mathcal C \\
 \mathcal C &= \text{cyclic rotation of tensor factors}. \nonumber
\end{align}
The entanglement entropy for any R\'enyi index $0 \le \alpha \le \infty$ is given by
\begin{align}
S
&= -\log \left[ \sum_{k=1}^n c_k \left( \frac{e^{2\pi i k /n}}{2} \right)^L \right] \nonumber\\
&= L \log 2 - \gcd(L,n) \log 2
\end{align}
where $c_k$ are multiplicities of the transfer matrix' eigenvalues.
We do not compute $c_k$ from the transfer matrix,
but they have to be determined by this formula
since $L \mapsto \gcd(L,n)$ is a periodic function.
The appearance of $\gcd$ is
because there are $\gcd(L,n)$ rings of cluster states.
This proves an interesting statement that
\begin{equation}
 c_k = \frac{1}{n} \sum_{j = 1}^{n} e^{2\pi i j k /n} 2^{\gcd(j,n)}
\end{equation}
are nonnegative integers for all $k = 1,2, \ldots, n$.

The examples of this section pose a challenge to unconditionally define the
subleading term $\gamma$.
For example, the following limit
\begin{equation}
\lim_{L \to \infty} S(L) - L( S(L+1) - S(L) )
\end{equation}
does not exist for the MPS state in Eq.~\eqref{eq:K}.
Even a Kitaev-Preskill-like combination,
canceling off the length (area) contribution, will not converge.
One might inevitably have to introduce some perturbation to define $\gamma$,
as our examples are not generic in the absence of any symmetry.

\section{Time reversal and lattice symmetries} \label{app:othersymmetries}

In this appendix,
we raise a question whether other symmetries on a 1D chain reduced from a 2D chain
can give rise to a robust finite sub-leading term of the entanglement entropy.
We consider three kinds of symmetries: the time reversal $\mathcal T$,
lattice reflection $\mathcal R$, and lattice inversion $\mathcal I$.
For 1D systems the lattice reflection and the inversion may coincide.
The reason we are distinguishing the two is that the 1D chain is divided into two parts, upper and lower, across the entanglement cut.
Upon the lattice reflection, the upper part and the lower part are not exchanged;
however, by the lattice inversion, which amounts to $\pi$-rotation about a point,
the two parts are exchanged.

The scope of this appendix is restricted to situations where the symmetry group $G$ is
$G = \mathcal T$, $G = \mathcal R$, $G = \mathcal I$, $G = \mathcal T \times \mathcal R$, or $G = \mathcal T \times \mathcal I$.
The argument below does not apply to a situation where $G = \mathcal R \times \mathcal I$  or $G = \mathcal T \times \mathcal R \times \mathcal I$.

We will find symmetry-respecting deformations of states,
after which the entanglement entropy becomes
\begin{equation}
 S_\text{new} = \alpha L
\end{equation}
for some $\alpha$, or
\begin{equation}
 S_\text{new} = \text{const.} \ge 0 .
\end{equation}
This will prove that if the non-positive subleading term ($-\gamma$) is robust under those symmetries, it must be zero.
This is in contrast to the situation where the 1D state form a nontrivial SPT under a product group of internal symmetry,
where a strictly negative subleading term ($-\gamma$) is stabilized by the internal symmetry.

\subsection{Lattice reflection}

Recall that our 1D chain has two physical qudits, $a_i$ and $b_i$, at each site $i$.
The entanglement cut separates the physical qudits
so that all $a_i$ qudits are in one partition and all $b_i$ are in another.
The lattice reflection is realized as
\begin{equation}
 \mathcal R : a_i \leftrightarrow a_{-i}, ~ b_i \leftrightarrow b_{-i}
\end{equation}
for all $i= \ldots, -2, -1, 0, 1, 2, \ldots$.

\begin{figure}[b!]
 \centering
 \includegraphics[width=0.4\textwidth]{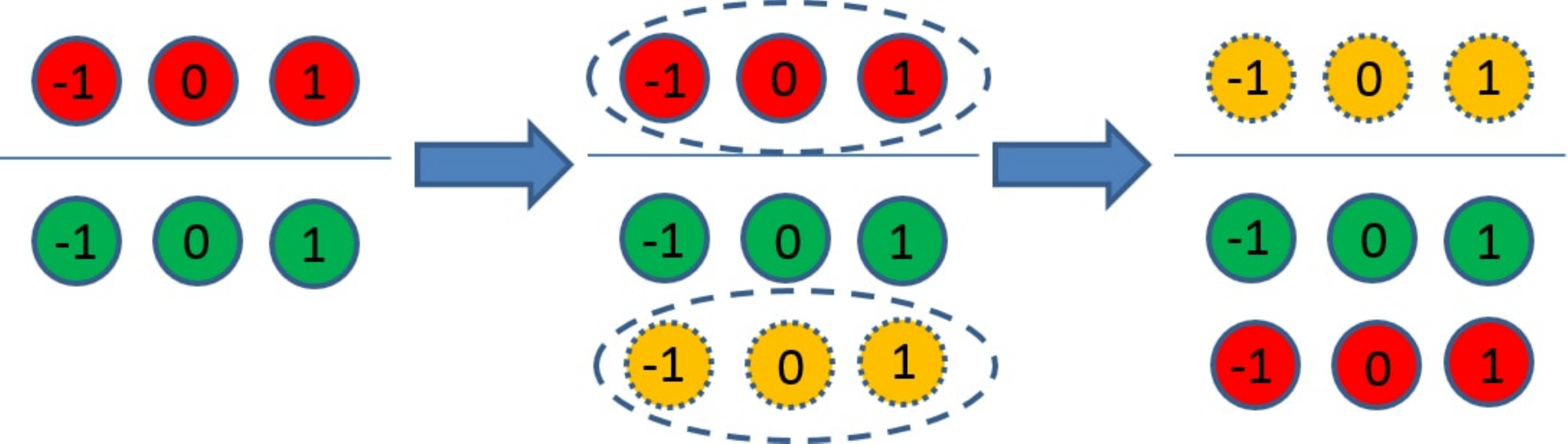}\\
 \begin{center}(a)\end{center}
 \includegraphics[width=0.4\textwidth]{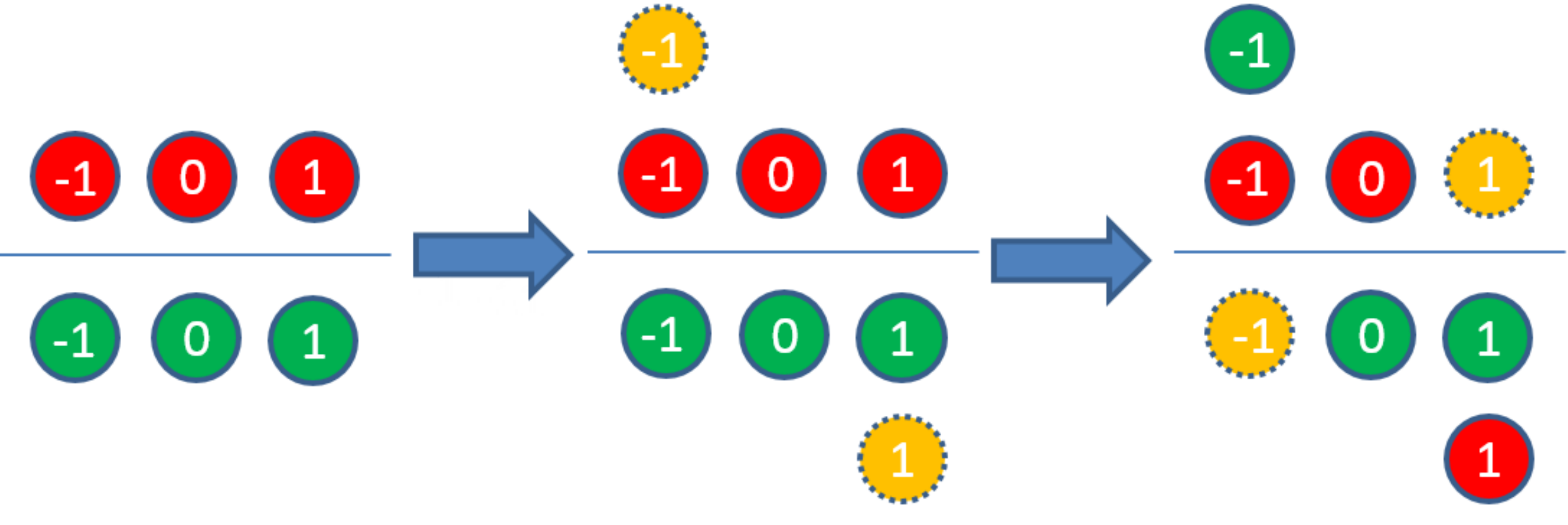}\\
 \begin{center}(b)\end{center}
 \caption{Deforming the state while respecting the lattice reflection or inversion symmetry.
  From the original 1D chain which consists of red ($a_i$) and green ($b_i$) qudits,
  one can insert an auxiliary yellow ($c_i$) qudit.
  Then one can apply the swap operator that exchanges the red qudits and yellow qudits circled by the dashed ellipses.
  This swap operation can be implemented continuously
  without breaking the lattice reflection symmetry.
  The numbers in each qudits label the positions of the corresponding qudits before the swap operation.
  It is understood that we have a 1D chain of qudits, although only a few sites are shown here.}
 \label{fig:lattice-symmetry}
\end{figure}

Consider inserting auxiliary qudits $c_i$ in the product state into the chain.
Each site is now consisting of $a_i, b_i, c_i$,
and we assume that $b_i, c_i$ belong to the same partition with respect to the entanglement cut.
Now we introduce the unitary operator $W_i$ on site $i$
that implements the swap between $a_i$ and $c_i$ (see Fig.~\ref{fig:lattice-symmetry}):
\begin{equation}
 W = \sum_{u,v} \ket{v,u} \bra{u,v}
 \label{eq:defSwap}
\end{equation}
The uniform application $\prod_i W_i$ obviously respects the lattice reflection symmetry $\mathcal R$.
Since $c_i$ were in the product state, they had no entanglement with the rest.
Thus, after the swap operation,
the entanglement entropy becomes identically zero across the existing cut.
One can implement $W$ smoothly since the unitary group is connected.
In this way, we have found a smooth deformation of the state
such that the entanglement entropy in the final state is simply zero.
In particular, we have smoothly changed the subleading term, if any, to zero.

\subsection{Lattice inversion without translation}

The lattice inversion is implemented as
\begin{equation}
 \mathcal I : a_i \leftrightarrow b_{-i}
\end{equation}
for all $i=\ldots, -1,0,1,\ldots$.
Similarly as in the previous subsection, by introducing auxiliary qudits in the product state and swap unitary,
one can push the physical qudits, expect those at $i=0$, to one side of the entanglement cut,
while respecting the lattice inversion symmetry (see Fig.\ref{fig:lattice-symmetry}).
The deformed state can be viewed as a 1D state where the entanglement cut divides
the chain into halves of length $L/2$.
The entanglement entropy does not depend on the system size
(the ``area'' law of entanglement entropy)
and is equal to some positive constant $h$.
If the chain was a nontrivial SPT under this inversion symmetry, such as the Haldane spin-1 chain,
then $h$ cannot be made to become zero;
if it was trivial SPT, then a smooth deformation such that $h \to 0$ is possible.

If we used Levin-Wen combination to define the subleading term ($-\gamma$),
then
\begin{equation}
 \gamma = S_{AB} + S_{BC} - S_{ABC} - S_B
\end{equation}
which is nonnegative by the strong subadditivity.
(At this point, we should not use the R\'enyi entropy, but von Neumann)
The deformed state clearly gives $\gamma =0$.

\subsection{Time reversal}

We assume that the system consists of spin-$J$'s, and the time reversal is implemented by
\begin{equation}
 \mathcal T = e^{-i\pi J_y} K
\end{equation}
for spin systems in the $J_z$-basis, where $K$ is the complex conjugation.
We will construct a similar deformation as in the previous cases.
We insert auxiliary spins in product states, and swap the spins of the original chain with the auxiliary ones.
Complication arises from two sources: The first one is that a half-integer spin cannot be time-reversal invariant.
This is easily resolved by inserting singlets formed by two spins.
The second one is that the swap $W$ and its smooth implementation $W(t)$ must commute with $\mathcal T$.
To resolve the second one, we will shortly prove that there exists $W_2(t)$ for any $J$ such that
\begin{align}
 &W_2(t=0) = I, \quad W_2(t=\pi) = W \otimes W,\nonumber\\
 &[W_2(t) , \mathcal T] = 0 \text{ for all } t.
\end{align}

\begin{figure}[tb]
 \includegraphics[width=.4\textwidth]{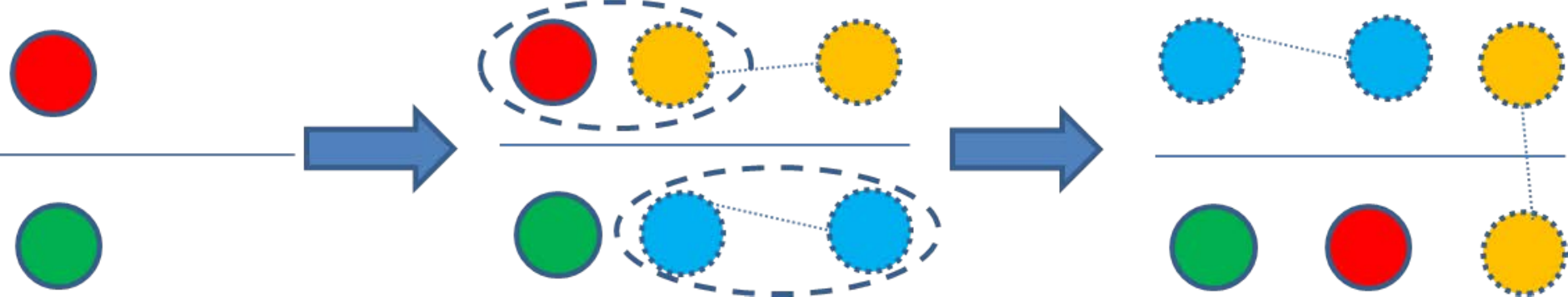}
 \caption{Deforming state while respecting symmetry.
 In addition to the red ($a_i$) and green ($b_i$) qudits, a pair of time reversal-invariant spin singlets are inserted in each site (the yellow qudits are $a_i'$ and blue qudits are $b_i'$).
 The swap unitary is applied to the qudits circled by the dashed ellipses, so that the entanglement across the cut solely arises from the inserted singlet.
 The swap can be implemented continuously during which the time-reversal symmetry is unbroken. It is understood we have a 1D chain of qudits, although only one site is shown here.
 }
 \label{fig:time-reversal}
\end{figure}

Equipped with $W_2=W_2(t)$,
we can deform the state so that the final state has entanglement entropy
\begin{equation}
 S = \alpha_J L
 \label{eq:ent-time}
\end{equation}
exactly without any subleading term, where $\alpha_J=\log(2J+1)$ is the entanglement entropy of a singlet consisting of two spin-$J$'s.
To see this, insert singlets $a_i'$ to the partition where $a_i$ belong, and another set of singlets $b_i'$ to the partition where $b_i$ belong.
Note that each of $a_i'$ or $b_i'$ consists of two spins, whereas each of $a_i$ or $b_i$ consists of one spin.
Apply $W_2$ such that the pair of $a_i$ and one auxiliary spin from $a_i'$ is exchanged with the whole singlet $b_i'$.
See Figure~\ref{fig:time-reversal}.
The original $b_i$ is not moved at all, and $a_i$ is brought to the partition where $b_i$ belongs.
The singlet $b_i'$ is moved to the opposite partition, and the singlet $a_i'$ is now shared between the entanglement partitions.
Thus, the entanglement entropy of the deformed state entirely comes from the singlets $a_i'$,
and Eq.~\eqref{eq:ent-time} holds.

Remark that this time-reversal invariant deformation respects lattice reflection and translation symmetry,
if they were present in the original state.
The deformation using $W_2$ can also be adapted to a situation where there is a lattice inversion symmetry.

We now construct the promised $W_2(t)$.
In the basis where $J_z$ is diagonal, we will show that there exists a real orthogonal matrix $W_2(t)$ such that
it commutes with $(e^{-i\pi J_y})^{\otimes 4}$, and it smoothly interpolates between the identity and $W \otimes W$.
Observe that $R = e^{-i \pi J_y}$ is a real matrix since $J_y=(J_+ - J_-)/2i$ is purely imaginary.
Since $J_y$ is hermitian, we have $R^T = R^\dagger = R^{-1}$.
Moreover, $R^2$ is $+1$ for integer spins or $-1$ for half-integer spins.
Therefore, $(R \otimes R)^2 = R^2 \otimes R^2 = 1$.
It follows that $R^{\otimes 2}$ is real symmetric with eigenvalues $\pm 1$.
The swap matrix $W$ is obviously real symmetric and squares to $1$.
Since $W$ and $R^{\otimes 2}$ commute, they can be simultaneously diagonalized
by a real orthogonal matrix.
Likewise, $W^{\otimes 2}$ and $R^{\otimes 4}$ can be simultaneously diagonalized,
and there exists a real orthonormal basis $\ket{w=\pm 1, r=\pm 1, k}$ consisting of common eigenvectors of $W^{\otimes 2}$ and $R^{\otimes 4}$,
where $w, r$ are the eigenvalue of $W^{\otimes 2}$ and $R^{\otimes 4}$, respectively,
and the index $k$ runs from 1 to the degeneracy $k_{w,r}$ of the common eigenspace.
We claim that both $k_{-1,+1}$ and $k_{-1,-1}$ are always even.
Given this claim, we can construct $W_2(t)$ by
\begin{align}
 W_2(t) \vert_{\mathrm{span}\{ \ket{-1,r,2m-1}, \ket{-1,r,2m} \}} =
 \begin{pmatrix}
  \cos t & \sin t \\ -\sin t & \cos t
 \end{pmatrix}
\end{align}
for $m = 1,\ldots, k_{-1,r}/2$, and the identity on $w=+1$ subspace.
The constructed $W_2(t)$ is clearly real orthogonal, and commutes with $R^{\otimes 4}$
since it preserves the eigenspaces of $R^{\otimes  4}$.
We have $W_2(0)=I$ by definition, and $W_2(\pi)=+1$ on the $w=+1$ subspace
and $W_2(\pi)=-1$ on the $w=-1$ subspace;
hence, $W_2(\pi) = W^{\otimes 2}$.

It remains to compute the degeneracy $k_{-1,r}$ to show that it is even.
Let $\{ \ket a : a = 1,\ldots, 2J+1 \}$ be a complete orthonormal set of eigenvectors of $R$.
These $\ket a$ may not be real vectors, but the degeneracy of eigenspaces can be computed
with respect to any basis we choose.
Then,
$(\ket{ab} \pm \ket{ba})/\sqrt{2}$ are complete common eigenvectors of $R \otimes R$ and the swap $W$.
Then, the $(-1)$-eigenvectors of $W^{\otimes 2}$
in an eigenspace of $R^{\otimes 4}$ are
\begin{equation}
 \frac{(\ket{ab}-\ket{ba})(\ket{cd}+\ket{dc})}{2}, ~ \frac{(\ket{ab}+\ket{ba})(\ket{cd}-\ket{dc})}{2}
\end{equation}
which always come in pairs.
The degeneracy is $k_{-1,r} = 2 \frac{N(N-1)}{2} \frac{N(N+1)}{2} = \frac12 N^2(N+1)(N-1)$,
an even number for any $N=2J+1$.

Without introducing $W_2$, a continuous real implementation of $W$ alone from the identity is not possible.
For spin-$\frac12$, the swap $W$ has determinant $-1$, so $W$ belongs to the non-identity component of $O(4)$.

\section{Cylinder extrapolation method on the 2D cluster state at nonzero temperature}
\label{app:thermal}

Here we calculate the mutual information for the cluster state
across the circumferential cut of a cylinder.
We consider the geometry of Fig.~\ref{fig:triangular-lattice}.
The mutual information
\begin{align}
I(A:B) = S(A) + S(B) - S(AB)
\end{align}
is preferred to the entropy $S(A)$ (or $S(B)$)
because it obeys an area law even at finite temperatures.%
\cite{CastelnovoChamon2007Entanglement,WolfVerstraeteHastingsEtAl2008Area}

The Hamiltonian is given by Eq.~\eqref{eq:clusterHamiltonian}.
By definition of the entropy,
the mutual information is invariant under local unitary in either region $A,B$,
and it is oblivious to any tensor product factor.
Hence, by the same argument as for the ground state of the cluster state,
the mutual information of the 2D cluster state
reduces to that of mutual information of the 1D cluster state with the extensive bipartition.
The Gibbs state at inverse temperature $\beta$ is given by
\begin{align}
\rho_{AB}
&= \mathcal Z ^{-1} \prod_{j}^{2L}
\exp\left( \beta \sigma^z_{j-1} \sigma^x_j \sigma^z_{j+1} \right)\\
&= \mathcal Z^{-1} \prod_{j}^{2L}
\left( I \cosh \beta + \sigma^z_{j-1} \sigma^x_j \sigma^z_{j+1} \sinh \beta \right).
\end{align}
The partition function $\mathcal Z$ is equal to that of uncoupled $2L$ spins in a magnetic field,
\begin{align}
\mathcal Z = 2^{2L} \cosh^{2L} \beta,
\end{align}
and the spectrum of $\rho_{AB}$ is the tensor product of $2L$ identical spectra
$\{ \frac12 (1 \pm \tanh \beta) \}$.
Hence, the von Neumann entropy is
\begin{align}
S(AB) = 2L f\left( \frac{1+t}{2} \right)
\end{align}
where $t = \tanh \beta$ and
$f(x) = -x \log x - (1-x) \log (1-x)$ is the binary entropy function.

Generalizing the result in Sec.~\ref{sec:FlatES},
we get the reduced density matrix for $A$
\begin{align}
\rho_A = \Tr_B (\rho_{AB}) = \frac{1}{2^{L}} \left(I + t^L \prod_{j \in A} \sigma^x_j \right),
\end{align}
whose entropy is
\begin{align}
S(A)
&= 2^{L-1} \left(\frac{1+t^L}{2^L} \log \frac{2^L}{1+t^L} + \frac{1-t^L}{2^L} \log \frac{2^L}{1-t^L} \right)\\
&= L \log 2 - \frac 12 \log(1-t^{2L}) - \frac{t^L}{2}\log\frac{1+t^L}{1-t^L} .
\end{align}
Therefore, the mutual information is
\begin{align}
I(A:B)
&= 2L(\log 2 - f((1+t)/2)) \nonumber \\
&\quad - \log(1-t^{2L}) - t^L\log\frac{1+t^L}{1-t^L} \\
&=
2\alpha_t L - t^{2L} + O(t^{4L})
\end{align}
where $\alpha_t = O(t^2)$ for small $t = \tanh \beta$ and $\alpha_t \sim \log 2$ for $t \sim 1$.

In conclusion, at any finite $\beta$, the subleading term is exponentially small in $L$.
Note that for any $0 < \beta \le \infty$,
the usual correlation length of the cluster state is zero
since it differ from a product Gibbs state by a quantum circuit of depth 2.
So, the length scale of the subleading term is greater than the usual correlation length,
and is diverging as $\beta \to \infty$.

\bibliography{gamma-ref}

\end{document}